\documentclass[useAMS,usenatbib]{mn2e}
\pdfoutput=1
\usepackage{txfonts}
\usepackage{natbib}
\usepackage{aas_macros}
\usepackage{graphicx}
\usepackage{subfigure}

\newcommand{\gsim}{\mathrel{\hbox{\rlap{\lower.55ex \hbox {$\sim$}}
                   \kern-.3em \raise.4ex \hbox{$>$}}}}
\newcommand{\lsim}{\mathrel{\hbox{\rlap{\lower.55ex \hbox {$\sim$}}
                   \kern-.3em \raise.4ex \hbox{$<$}}}}

\title[Gas accretion onto planetary cores]{Gas accretion onto planetary cores: three-dimensional self-gravitating radiation hydrodynamical calculations}
\author[B.A. Ayliffe \& M.R. Bate]{Ben A. Ayliffe\thanks{E-mail:
ayliffe@astro.ex.ac.uk} and Matthew R. Bate\thanks{E-mail:
mbate@astro.ex.ac.uk}\\ School of Physics, University of Exeter, Stocker
Road, Exeter EX4 4QL}

\date{\today}
\begin{document}
\maketitle
\begin{abstract}
We present results from three-dimensional, self-gravitating radiation hydrodynamical models of gas accretion by planetary cores.  In some cases, the accretion flow is resolved down to the surface of the solid core -- the first time such simulations have been performed. We investigate the dependence of the gas accretion rate upon the planetary core mass, and the surface density and opacity of the encompassing protoplanetary disc. Accretion of planetesimals is neglected.

We find that high-mass protoplanets are surrounded by thick circumplanetary discs during their gas accretion phase but, contrary to locally-isothermal calculations, discs do not form around accreting protoplanets with masses $\lsim 50$~M$_{\earth}$ when radiation hydrodynamical simulations are performed, even if the grain opacity is reduced from interstellar values by a factor of 100.

We find that the opacity of the gas plays a large role in determining the accretion rates for low-mass planetary cores.  For example, reducing the opacities from interstellar values by a factor of 100 leads to roughly an order of magnitude increase in the accretion rates for $10-20~\rm M_{\earth}$ protoplanets. The dependence on opacity becomes less important in determining the accretion rate for more massive cores where gravity dominates the effects of thermal support and the protoplanet is essentially accreting at the runaway rate. Increasing the core mass from 10~$\rm M_{\earth}$ to 100~$\rm M_{\earth}$ increases the accretion rate by a factor of $\approx 50$ for interstellar opacities. Beyond $\rm \sim 100~M_{\earth}$ the ability of the protoplanetary disc to supply material to the accreting protoplanet limits the accretion rate, independent of the opacity.

Finally, for low-mass planetary cores ($\rm \lsim 20M_{\earth}$), we obtain accretion rates that are in agreement with previous one-dimensional quasi-static models.  This indicates that three-dimensional hydrodynamical effects may not significantly alter the gas accretion timescales that have been obtained from quasi-static models. 

\end{abstract}

\section{Introduction}

The favoured theory of gas giant planet formation is described by the core accretion model \citep{perri74,mizuno78}. This bottom-up formation mechanism begins with the coagulation of grains immersed in a protoplanetary disc to form planetesimals. These planetesimals are then further agglomerated to form large solid cores. The fate of these cores is dependent upon their mass and surroundings. The lower mass cores will likely become terrestrial planets. High mass cores ($\gtrsim 10~M_{\earth}$) that form at a time and in a region of plentiful gas are expected to accrete large gas atmospheres, such as we see for Jupiter and Saturn. Whilst those high mass cores that form later, perhaps when the nebula is dissipating, are more likely to resemble Neptune and Uranus, with comparatively small atmospheres.  See \cite{pandp5} for a recent review of the core accretion model.

A major challenge faced by proponents of the core accretion model is to explain how the entire process of forming a Jupiter-like planet might occur within the relatively short-lived protoplanetary discs that are observed. Observations of dust in discs around young stars suggest disc lifetimes of $\lsim 10$~Myrs \citep[e.g.][]{haisch2001}.  A large fraction of the overall formation time may be used in the formation of the core itself. Core growth using the conditions of \cite{lissauer87}, which include a nebula density several times that of the minimum mass solar nebula \citep[MMSN,][]{hayashi81} and a high constant rate of planetesimal accretion, allowed \cite{pollack96} to assemble high mass cores ($\rm \gtrsim 10~M_{\earth}$) in around $6 \times 10^{5}$ years. \cite{fortier2007} considered oligarchic growth of solid cores, and found that a 42~$M_{\earth}$ core might be formed in as little as 1.1~Myrs with a nebula density of 10 times the MMSN, and the inclusion of atmospheric gas drag on relatively small planetesimals ($\approx 10$~km radius). However, they found that at a nebula density equivalent to the MMSN case they were unable to form a 10~$\rm M_{\earth}$ core in less than 10~Myrs. The core masses of giant planets in our solar system are still not known with certainty. Interior models by \cite{guillot2004} put Jupiter's core mass in the range $0-11~{\rm M}_{\earth}$, whilst Saturn's is constrained to $\rm 9-22 M_{\earth}$. Radial migration of the protoplanet can help to shorten the formation timescale by providing the protoplanet access to regions of the disc that are undepleted in planetesimals  (\citealt*{AliMorBen2004}; \citealt{Alibertetal2005}).  Though the conditions and processes by which solid cores form may not yet be fully understood, the range of allowed core masses (particularly for Saturn) suggest that form they must.

Assuming that massive cores are able to form within a disc's lifetime, the next evolutionary stage is one of gas accretion to form a massive atmosphere. A high mass solid core will possess a small gas atmosphere that was concurrently accreted with the bulk of the solid material \citep{pollack96}. However, the majority of gas accretion is thought to take place during a runaway gas accretion phase that begins when the atmosphere has a mass approximately equal to that of the solid core (the so-called crossover mass). The intervening period is thought to be one of steady gas and planetesimal accretion prior to reaching the crossover mass \citep[so-called Phase 2 accretion:][]{pollack96,hubickyj2005}. The factors that affect the gas accretion rate include the mass of the core, the rate of planetesimal bombardment (which deposits thermal energy in the protoplanet's gaseous envelope), and the opacity, temperature, and density of the nebula.

Recently, \cite{hubickyj2005} examined the effects of grain opacity using one-dimensional quasi-static models for the envelopes around their evolving protoplanets. They use both interstellar grain opacities (IGO) and opacities of 2\% this level.  The justification for reduced grain opacities in the envelopes of forming protoplanets comes from calculations by \cite{podolak2003}.  These calculations showed that small grains, which most effectively absorb or scatter radiation, tend to remain suspended in the envelope for longer periods, consequently growing larger in size and so losing their aforementioned effectiveness.  Grains that do sediment to regions of higher temperature sublimate (above about 1500K for silicate grains). This sublimation disproportionately affects small grains, once again removing the most effective scatterers/absorbers, and so reducing the grain opacity that might be expected in the envelope surrounding a planetary core. Finally, given that, by definition, the protoplanetary disc has produced solid cores, the typical protoplanetary disc grains are likely to be larger than interstellar grains, and thus less effective in providing opacity.   \cite{hubickyj2005} find at that 2\% IGO the time spent in Phase 2 of the accretion process is shortened by 70\% for their baseline case. \cite{papa2005} also consider the implications of reduced grain opacities on the gas accretion rate, and particularly Phase 2. Using a 1\% IGO they find up to two orders of magnitude reduction in the length of Phase 2, compared with their standard opacity case.

Despite the fact that gas accretion by a protoplanet from a protoplanetary disc is an intrinsically three-dimensional, hydrodynamical problem, the past calculations mentioned above that studied the evolution of a protoplanet have treated the problem using one-dimensional, quasi-static models.  Three-dimensional hydrodynamical calculations examining the interaction of a protoplanet with a disc to study planet migration have usually assumed a locally isothermal equation of state, and/or do not resolve the gas flow far inside the planet's Hill radius (\citealt{Brydenetal1999,Kley1999}; \citealt*{LubSeiArt1999}; \citealt{Nelsonetal2000,Masset2002}; \citealt{batezeus}; \citealt*{DAnHenKle2003}; \citealt{NelPap2004}; \citealt*{PapNelSne2004}; \citealt{DAnBatLub2005}; \citealt{LubDAn2006,PaaMel2006}; \citealt*{MasDAnKle2006, DAnLubBat2006}; \citealt{DAnLub2008}).  Recently, some hydrodynamical calculations have departed from the assumption of local isothermality, investigating the effects of different equations of state and/or radiation transport \citep*{DAnHenKle2003, MorTan2003, KlaKle2006, PaaMel2006, PaaMel2008}, while \cite{FouMay2008} have performed calculations including both radiative transfer and self-gravity.  These papers report that the inclusion of non-isothermal physics generally results in reduced, or even outward, radial migration rates.  However, none of these calculations have resolved the flow much inside the Hill radius of the protoplanet, limiting their ability to examine the gas accretion rates of the protoplanets.

In this paper, we perform three-dimensional, self-gravitating, radiation hydrodynamical calculations of gas accretion by protoplanets with masses ranging from $10-333$ M$_{\earth}$.  Some of the calculations resolve the accretion flow down to the surface of the solid planetary core.  Our aims are to investigate whether and how the inclusion of three-dimensional hydrodynamics changes the conclusions obtained from previous one-dimensional quasi-static models of gas accretion and to examine how the inclusion of radiative transfer alters the accretion rates obtained from three-dimensional locally-isothermal calculations.  The models presented here neglect planetesimal bombardment and radial migration of the protoplanet. We aim to examine the effect of these processes in a future paper.  In this paper, we investigate the dependence of the gas accretion rates on the mass of the protoplanet, the surface density of the protoplantary disc, and the opacity of the gas.

In Section 2, we describe our computational method.  In Section 3, we examine the effects of changing the model parameters (e.g. resolution) on the results.  Section 4 presents the results from both our locally-isothermal and self-gravitating radiation hydrodynamical simulations, while in Section 5 we discuss the implications of our results for giant planet formation theory. Our conclusions are given in Section 6.

\section{Computational method}
\label{method}

The calculations described herein have been performed using a three-dimensional smoothed particle hydrodynamics (SPH) code. This SPH code has its origins in a version first developed by \citeauthor{benz90} (\citeyear{benz90}; \citealt{benzcam90}) but it has undergone substantial modification in the subsequent years. Energy and entropy are conserved to timestepping accuracy by use of the variable smoothing length formalism of \cite{springel2002} and \cite{Monaghan2002} with our specific implementation being described in \cite{price2007}. Gravitational forces are calculated and neighbouring particles are found using a binary tree. Radiative transfer is modelled in the flux-limited diffusion approximation using the method developed by \citet{WhiBatMon2005} and \citet{WhiBat2006}.  Integration of the SPH equations is achieved using a second-order Runge-Kutta-Fehlberg integrator with particles having individual timesteps \citep*{bate95}. The standard implementation of artificial viscosity is used \citep{mongin83,mon92} with the parameters $\alpha_{v}$ = 1 and $\beta_{v}$ = 2. The code has been parallelised by M. Bate using OpenMP.

\subsection{Model setup}
\label{sec:model}

We perform our calculations in the reference frame of the planet, which orbits a star of mass $M_{*}$ at radius $r_{\rm p}$ with an angular speed given by $\Omega_{p} = \sqrt{GM_{*}/r_{p}^{3}}$, neglecting the mass of the planet.  We take the stellar mass to be 1$~\rm M_{\sun}$ and the orbital radius of the planet to be 5.2 AU.  Our standard protoplanetary disc has the same parameters as used by \citet{LubSeiArt1999} and \citet{batezeus}.  The initial radial temperature profile for the disc is $T_{\rm g} \propto r^{-1}$, which leads to a constant ratio of disc scaleheight with radius of $H/r=0.05$. The initial surface density of the disc has as $\Sigma \propto r^{-1/2}$ profile, with a value of 75~$\rm g~cm^{-2}$ at the planet's orbital radius in our standard model.

\begin{figure}
\centering
\includegraphics[width=8cm]{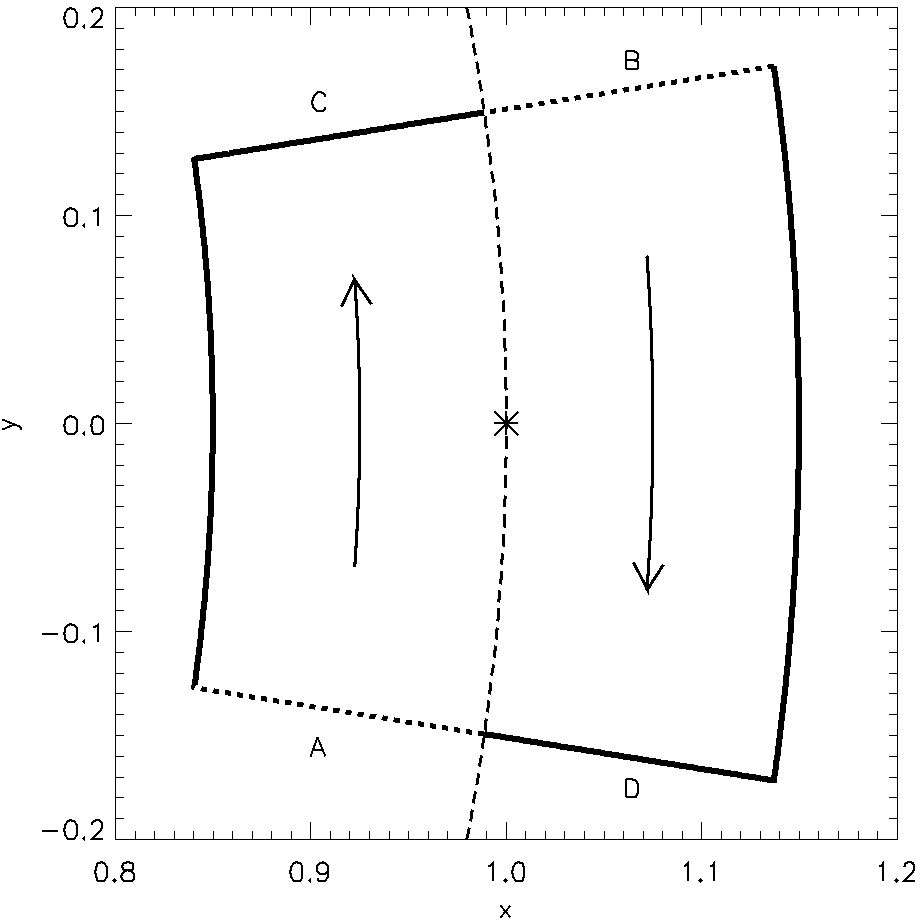}
\caption{The section of disc that we model with the location of the protoplanet marked by an asterisk. The section is fixed in the rotating reference frame of the protoplanet, such that the dashed line curving through its centre is the line of zero azimuthal velocity relative to the protoplanet. The dotted lines on the azimuthal boundaries (denoted A and B) mark the edges at which particles are injected to simulate gas flow past the planet. These particles are either captured by the protoplanet or leave the simulation region primarily through the opposite edges (C and D, respectively).}
\label{fig:secfig}
\end{figure}

In order to achieve high resolution near a protoplanet while maintaining a reasonable computation time we model only a small section of the protoplanetary disc centred on the planetary core as illustrated in Figure \ref{fig:secfig}. Our standard section size is $r=1\pm 0.15~r_{\rm p}$ ($5.2 \pm 0.78~{\rm AU}$), and $\phi = \pm 0.15$ radians.

\begin{figure*}
\centering
\includegraphics[width=16cm]{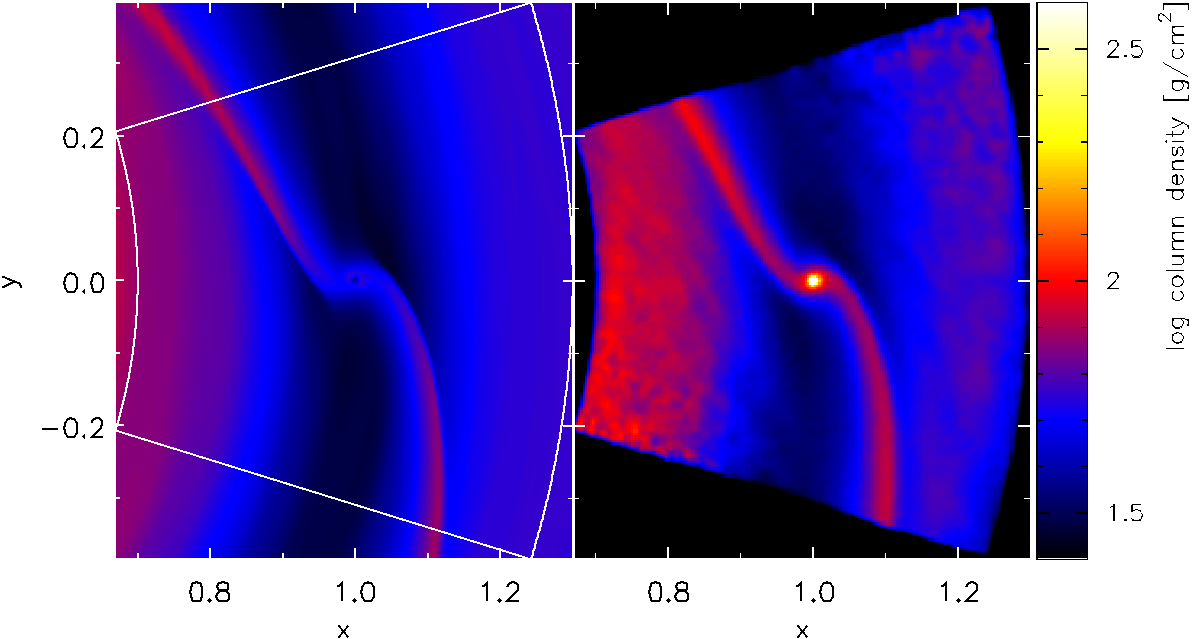}
\caption{Surface density plots of our disc region for a $\rm 33~M_{\earth}$ protoplanet using a locally-isothermal equation of state. The left panel is a zoomed view of the original ZEUS calculation of \protect \cite{batezeus} with the boundary of the section that we model overlaid as a white line. The right panel shows an equivalent SPH simulation modelling only the disc section and using SPH particle injection derived from the model shown in the left panel. The ZEUS calculation accretes gas that comes within 20\% of the protoplanet's Hill radius, while in the SPH calculation the protoplanet accumulates a gaseous envelope. The spiral shocks and gap in the protoplanetary disc generated by the protoplanet are well reproduced in the SPH model.}
\label{fig:zeuscomp}
\end{figure*}

\subsection{Boundary conditions}

The distribution of particles within the disc section is initially that of an unperturbed disc, with Keplerian velocities. Without further injection of particles along edges A and B of the disc section in Figure \ref{fig:secfig}, these gas particles would either be captured by the protoplanet or move outside of the boundaries of the disc section (typically through edges C and D, respectively) and be removed from the simulation.  To maintain the gas within the disc section and replicate the expected distribution for a disc about a planet (i.e., with a gap) we inject particles along edges A and B in Figure \ref{fig:secfig} whose density and velocity distributions are taken from global calculations of protoplanets embedded in protoplanetary discs. The global calculations we use are those of \cite{batezeus} who performed three-dimensional global simulations using a version of the ZEUS code \citep{stone}. \citeauthor{batezeus} performed calculations for protoplanets with masses of 1, 3, 10, 33, 100, and 333 M$_{\earth}$.  For our SPH calculations that have protoplanet core masses differing from these values we interpolate between the density and velocity distributions available from the ZEUS calculations.  After a few orbits the initial particle distribution is entirely lost and is replaced with the particle distribution due to the injected particles, creating the gap and other global features that are expected to be found when a protoplanet is embedded within a gaseous disc. 
Ghost particles are employed along the boundaries of the disc section to provide the viscous and pressure forces that would come from the gas which is not modelled outside of the boundaries.

The left panel of Figure \ref{fig:zeuscomp} shows a surface density plot from the ZEUS calculations of \cite{batezeus} for a 33~$\rm M_{\earth}$ protoplanet embedded in a locally isothermal disc. The right panel of the same figure shows the surface density plot from the equivalent SPH calculation (modelling only our standard section of the disc). The gap and spiral density waves in the disc are well reproduced in the SPH calculation.  The most obvious difference between the two plots is at the core. The ZEUS calculation treats accretion by removing gas that falls within 20\% of the core's Hill radius. By removing gas near the core the ZEUS models do not show the increased density expected with the accumulation of gas. This SPH model instead allows gas to accumulate on the surface of the protoplanet which, in this case, is positioned at 5\% of the Hill radius, $R_{\rm H}$.  Thus, the region near the protoplanet has a much higher surface density in the SPH calculation than in its ZEUS counterpart. The gridding used in the ZEUS calculation also leads to low resolution in the region surrounding 0.2 $\rm R_{H}$ at which gas is removed, leading to poor resolution near this boundary. In contrast, the resolution in the SPH calculation inherently increases where particles accumulate, allowing the flow near the core to be followed in detail.

The flux-limited diffusion scheme transfers energy between SPH particles, thus rendering it unable to radiate into a vacuum (where there are no particles). To allow radiation to escape from the surfaces of the disc we employ surface boundary regions that maintain the initial temperature profile in the high atmosphere of the disc. This boundary is situated above and below the midplane where the optical depth into the disc is $\tau \approx 1$. SPH particles within the boundary regions are evolved normally, except that they interact with the SPH particles in the bulk of the disc without their temperatures being affected.

\subsection{Equation of state}

Two equations of state are used for the calculations presented in this paper.  The first is a locally-isothermal equation of state as used in the ZEUS calculations of \cite{batezeus} with the temperature of the gas throughout the disc being a fixed function of radius as described in Section \ref{sec:model}. 

In the second case, radiation hydrodynamical calculations are conducted using an ideal gas equation of state $p=\rho T_{g} R_{g}/\mu$ where $R_{g}$ is the gas constant, $\rho$ is the density, $T_{g}$ is the gas temperature, and $\mu$ is the mean molecular mass. The equation of state takes into account the translational, rotational, and vibrational degrees of freedom of molecular hydrogen (assuming an equilibrium mix of para- and ortho-hydrogen; see \citealt{BlaBod1975}).  It also includes the dissociation of molecular hydrogen, and the ionisations of hydrogen and helium.  The hydrogen and helium mass fractions are $X=0.70$ and $Y=0.28$, respectively, whilst the contribution of metals to the equation of state is neglected.  More details on the implementation of the equation of state can be found in \cite{WhiBat2006}.  We note that the correction to the equation of state pointed out by \cite{Boleyetal2007} has been made. 

The two temperature (gas and radiation) radiative transfer in the flux-limited diffusion approximation employed in this work is implemented as described by \cite{WhiBatMon2005} and \cite{WhiBat2006}. Briefly, work and artificial viscosity (which includes both bulk and shear components) increase the thermal energy of the gas, and work done on the radiation field increases the radiative energy which can be transported via flux-limited diffusion. The energy transfer between the gas and radiation fields is dependent upon their relative temperatures, the gas density, and the gas opacity.

\begin{figure}
\centering
\includegraphics[width=8cm]{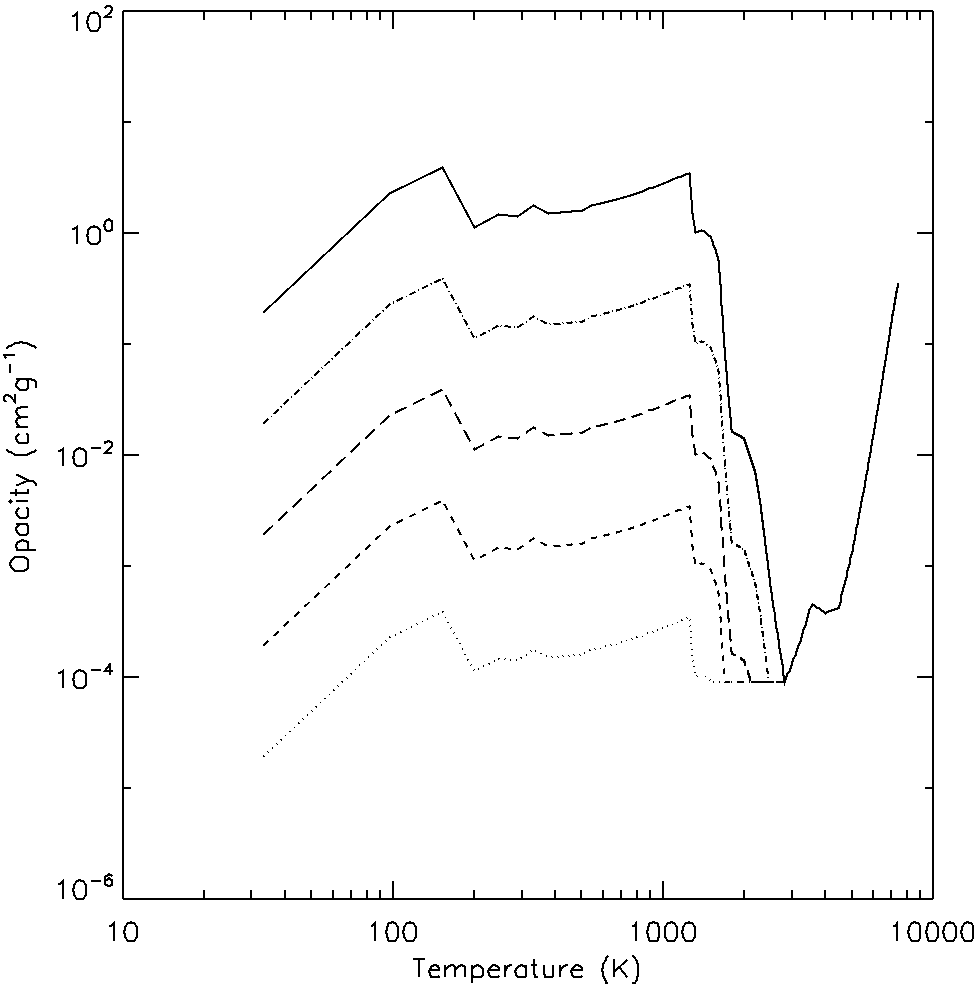}
\caption{Opacity as a function of temperature for gas of density $\sim 4.0 \times 10^{-11} {\rm g~cm}^{-3}$ over the temperature range $30-7200$~K, with different scaling factors: standard interstellar grain opacities (IGO) (solid), 10\% IGO (dot-dash), 1\% IGO (long-dash), and 0.1\% IGO (short-dash). An extreme case of 0.01\% IGO (dotted) is included to demonstrate the treatment at the low temperature end of the scale.}
\label{fig:kap}
\end{figure}

\begin{figure*}
\centering

\subfigure 
{
	\includegraphics[width=8cm]{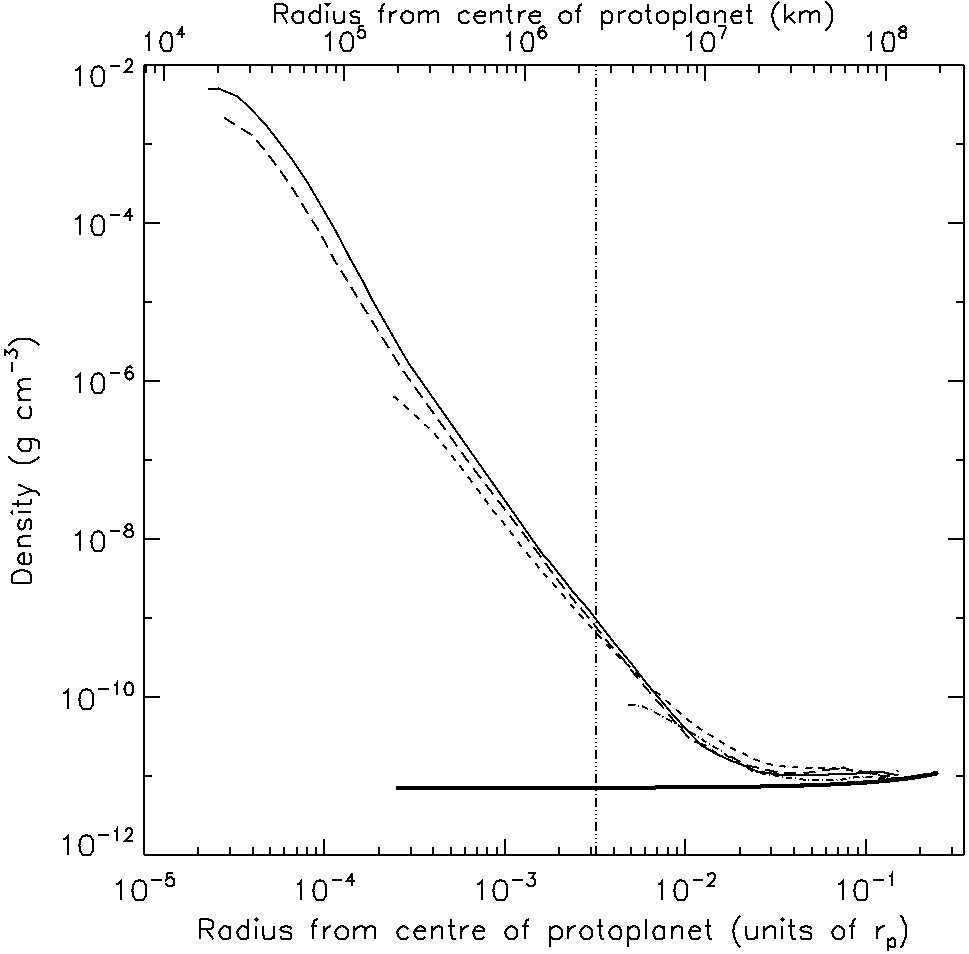}
}
\hfill
\subfigure 
{
	\includegraphics[width=7.8cm]{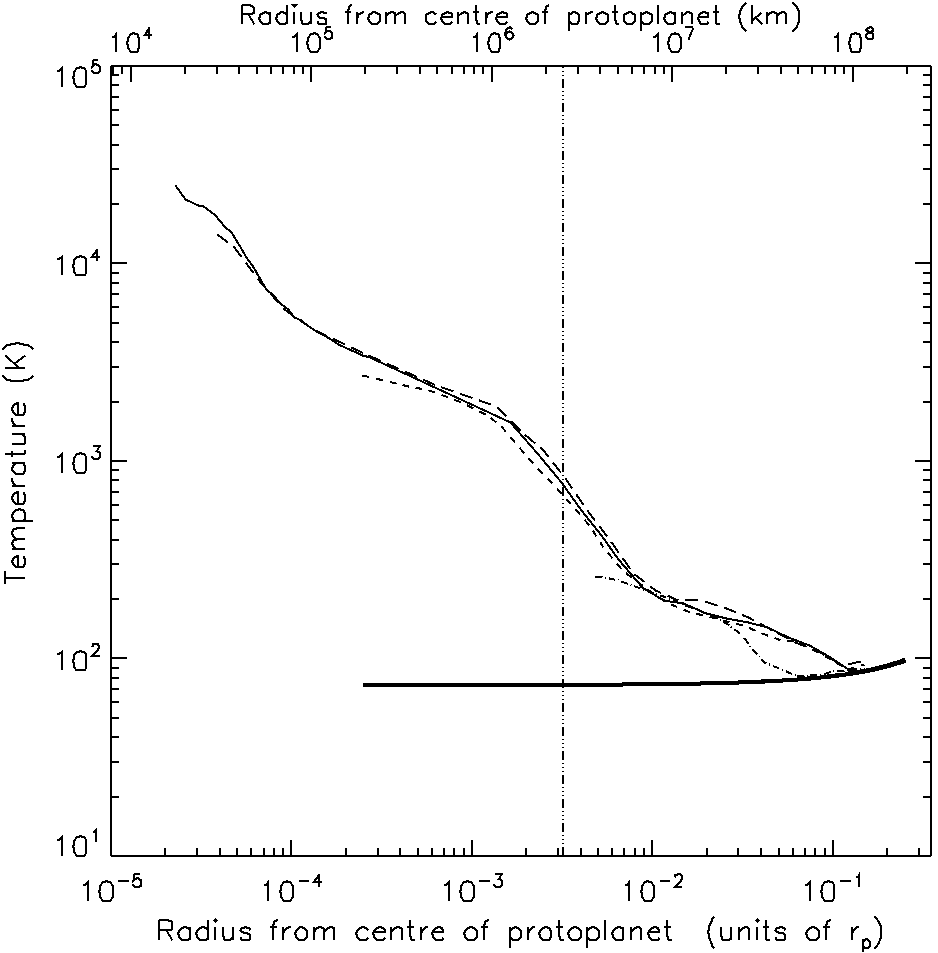}
}

\caption{The panels show the evolution of the radial density and temperature distributions of the protoplanetary envelope at different stages within 10 orbits. The radial distance is measured away from a 33~M$_{\earth}$ solid core towards the central star. The times shown are 0.1, 0.5, 1, and 10 orbits represented by dot-dash, short-dash, long-dash, and solid lines respectively. The thick solid line in both cases shows the initial conditions employed in the disc. Once the core has shrunk smoothly to its final radius the distributions change only as the envelope accretes more mass, and the form of the distribution remains similar.}
\label{fig:timeevolution}
\end{figure*}

\subsection{Opacity treatment}

We use interpolation of the opacity tables of \cite{pollack85} to provide the interstellar grain opacities (IGO) for solar metallicity molecular gas, whilst at higher temperatures when the grains have sublimated we use the tables of \cite{alexander75} (the IVa King model) to provide the gas opacities (for further details see \citealt{WhiBat2006}).

The agglomeration of interstellar grains into larger grains, or the sublimation of small grains, may lead to grain opacities that are lower than the IGO.  We mimic these effects by scaling the IGO values of \cite{pollack85} by various factors.  Note that the gas opacities from \cite{alexander75} that we use beyond the grain sublimation point are not scaled (the gas is still assumed to have solar metallicity). Figure \ref{fig:kap} plots opacity versus temperature for a given gas density with different scaling factors.  Between the temperatures corresponding to the peak grain opacity (occurring at $\approx 1500$~K) and the total sublimation of the dust grains, the reduced opacity may be decreased at most to the value at the latter temperature (note the flat section near the sublimation point in figure \ref{fig:kap} for low grain opacity cases). Opacities at temperatures lower than that of the peak grain opacity have no enforced minimum value; the 0.01\% IGO case demonstrates this case with the reduced opacity at 40~K having a lower value than at $\approx 2000$~K.

\subsection{Gas accretion rates}

Within this paper, gas accretion rates are calculated as the rate at which gas passes into the self-consistently calculated Hill sphere of the protoplanet given by
\begin{equation}
R_{\rm H}=\sqrt[3]{\frac{M_{\rm H}}{3M_{*}}}r_{\rm p}
\label{eq:hill}
\end{equation}
where $M_{\rm H}$ is the mass contained within $R_{\rm H}$ (i.e. the mass of the protoplanet core $M_{\rm p}$ and any gas within $R_{\rm H}$).  The accretion rate is averaged over the last orbit for each calculation once the simulation has reached 10 orbits. Accretion rates measured at identical times for differing physical conditions (e.g., planetary core masses and grain opacities) allow us to quantify the effects of changes in these parameters.

\begin{table}
\centering
\begin{tabular}{c c c c c}
\hline\hline
Core mass & Core mass & Hill radius &  Hill radius &  Solid core radius \\ 
$[{\rm M}_{\odot}]$ & $[{\rm M}_{\earth}]$ & [${\rm r}_{\rm p}$] &  [$\rm R_{\earth}$] &  [$\rm R_{\earth}$] \\ 
\hline
$3 \times 10^{-5}$  &10 & 0.022 & 2600 & 2.4   \\
$7 \times 10^{-5}$  &22 & 0.027 & 3300 & 2.7   \\
0.00010  &33 & 0.032 & 3900 & 3.1    \\
0.00017  &56 & 0.038 & 4600 & 3.5  \\
0.0003 & 100 & 0.046 & 5700 & 3.9  \\
0.0005 & 166 & 0.055 & 6700 & 4.1  \\
0.001 & 333 & 0.069 & 8400 & 4.4  \\ [1ex]
\hline
\end{tabular}
\caption{The protoplanet masses that we consider in this paper (in solar masses and Earth masses), along with their Hill radii (in units of the protoplanet's orbital radius and Earth radii) and their realistic physical solid core radii (based on the models of \citealt{seager07}).}
\label{table:radii}
\end{table}

\subsection{Planetary Cores}

The planetary cores in these simulations are modelled by a gravitational potential, and a surface potential that yields an opposing force upon gas within one core radius of the core's surface. The combination of the gravitational and surface forces takes the form of a modification to the usual gravitational force as
\begin{equation}
F_{r} = - \frac{GM_{p}}{r^{2}}\left(1 - \left(\frac{2R_{p}-r}{R_{p}}\right)^{4}\right)
\label{eq:surface}
\end{equation}
for $r <2~R_{\rm p}$ where $r$ is the radius from the centre of the planetary core, $R_{\rm p}$ is the radius of the core, and $M_{\rm p}$ is the mass of the core.  This equation yields zero net force between a particle and the planetary core at the surface radius $R_{\rm p}$, whilst inside of the core's radius the force is outwards and increases rapidly with decreasing radius. Gas particles therefore come to rest very close to the core radius, though the equilibrium position is slightly inward of this value due to the pressure exerted by the gas that accumulates on top of the inner most layer of particles.

The majority of our calculations use cores with radii equivalent to 1\% of their respective Hill radii. This is 10-20 times larger than the radius of a physical solid core (Table \ref{table:radii}).  However, since it is still much smaller than the Hill radius (the radius where accretion from the protoplanetary disc to the protoplanet is expected to occur), it is not expected that making the core radius smaller than this should greatly change the accretion rate. The benefit is that the calculations use much less compute time than if realistic core sizes are used.

However, for some of the calculations we do model the accretion process all the way down to the surface of realistically-sized planetary cores.  In these cases, the radii of the solid cores are taken from the models by \cite{seager07} for solid exoplanets comprised of 75\% water, 22\% silicates, and 3\% iron. Table \ref{table:radii} lists the Hill radii and realistic solid core radii for each of the different core masses that we consider.

\begin{figure}
\centering
\includegraphics[width=8cm]{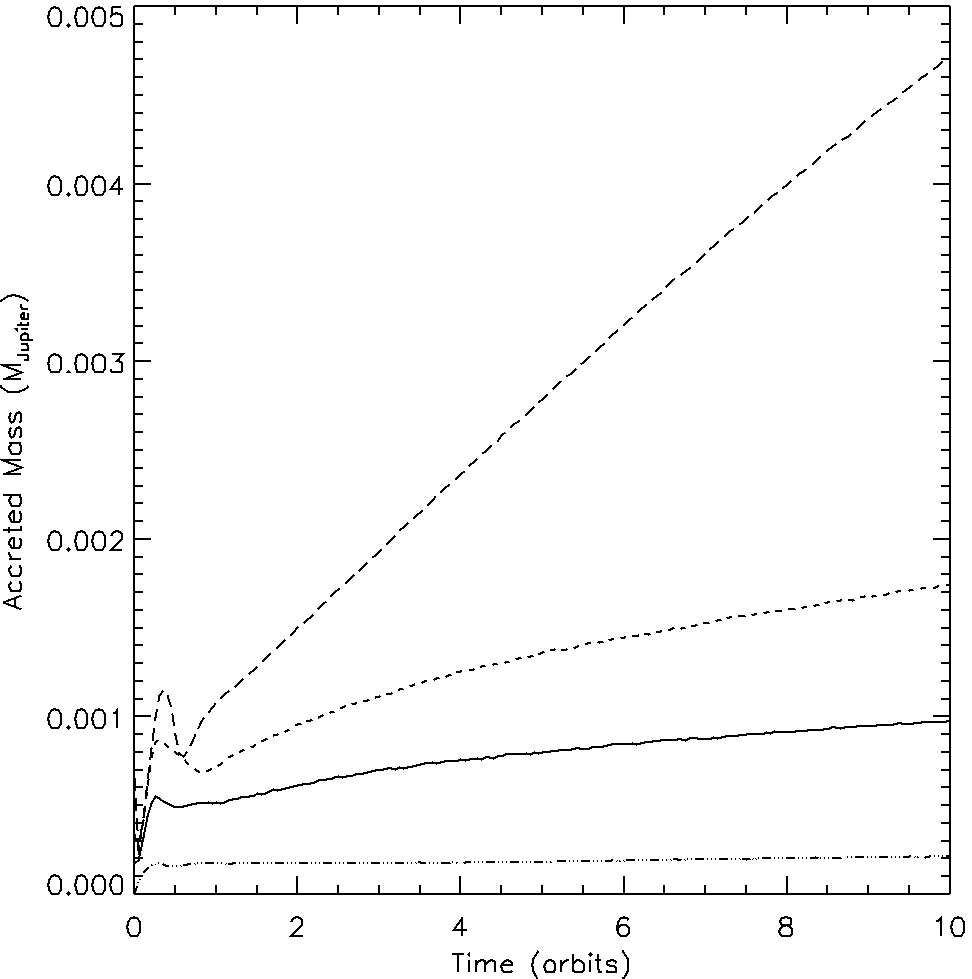}
\caption{The increasing gas mass within the Hill radius over the first 10 orbits is shown for four different core masses; 10~$\rm M_{\earth}$ (dash-dots-dash), 33~$\rm M_{\earth}$ (solid), 56~$\rm M_{\earth}$ (short-dash), and 100~$\rm M_{\earth}$ (long-dash). We measure our accretion rates over the last orbit.  These calculations were performed with our standard protoplanetary disc surface density (75~g~cm$^{-2}$) and standard interstellar grain opacities.}
\label{fig:accretionplots}
\end{figure}

To provide a smooth start to the radiation hydrodynamical calculations, $R_{\rm p}$ is initially set to 0.01~$r_{\rm p}$ and reduces exponentially in size to the desired radius (either 1\% of the Hill radius, or the physical core radius) during the first orbit of the protoplanet. Figure \ref{fig:timeevolution} illustrates the evolution of the gas density and temperature distribution along the radial axis towards the star at four different times for our standard case.  Our standard case is a 33~$\rm M_{\earth}$ planetary core embedded in a protoplanetary disc with an unperturbed surface density at $r_{\rm p}$ of 75~g~cm$^{-2}$ with standard interstellar grain opacities.  After the first orbit, the distributions change very little, slowly extending to greater values of density and temperature only as more gas is accreted.  Figure \ref{fig:accretionplots} shows the increase of the mass contained within the Hill sphere over the first ten orbits for several different core masses.  The settling to a quasi-equilibrium accretion rate can be seen to take less than 2 orbits.

With radiative transfer, the protoplanetary disc develops a vertical temperature structure even at large distances from the protoplanet due to viscous heating. However, at the boundaries of our disc section the departure of temperatures from those in the locally-isothermal model (or the temperature in the atmosphere of the disc) is very small.  This is apparent in Figure \ref{fig:timeevolution} from the convergence of the temperature distributions to those of the initial temperature distribution (thick solid line) at large radii from the protoplanet.  Typically, the dense gas at the midplane, far from the planet (i.e. where the disc is relatively unperturbed), is around 40\% hotter than that in the locally-isothermal model upon which our particle injection method is based.  One might question whether or not this slight mismatch at the boundary leads to noticeable effects.  However, this departure from isothermality is small compared with the temperature changes due to the accretion processes occurring at the planet.  Furthermore, in Section \ref{sec:secsize} where we discuss the impact of varying our radial and azimuthal boundaries, we find that our results are unaffected by moving the injection boundary further from the planetary core.

\begin{figure*}
\centering
\includegraphics[width=16cm]{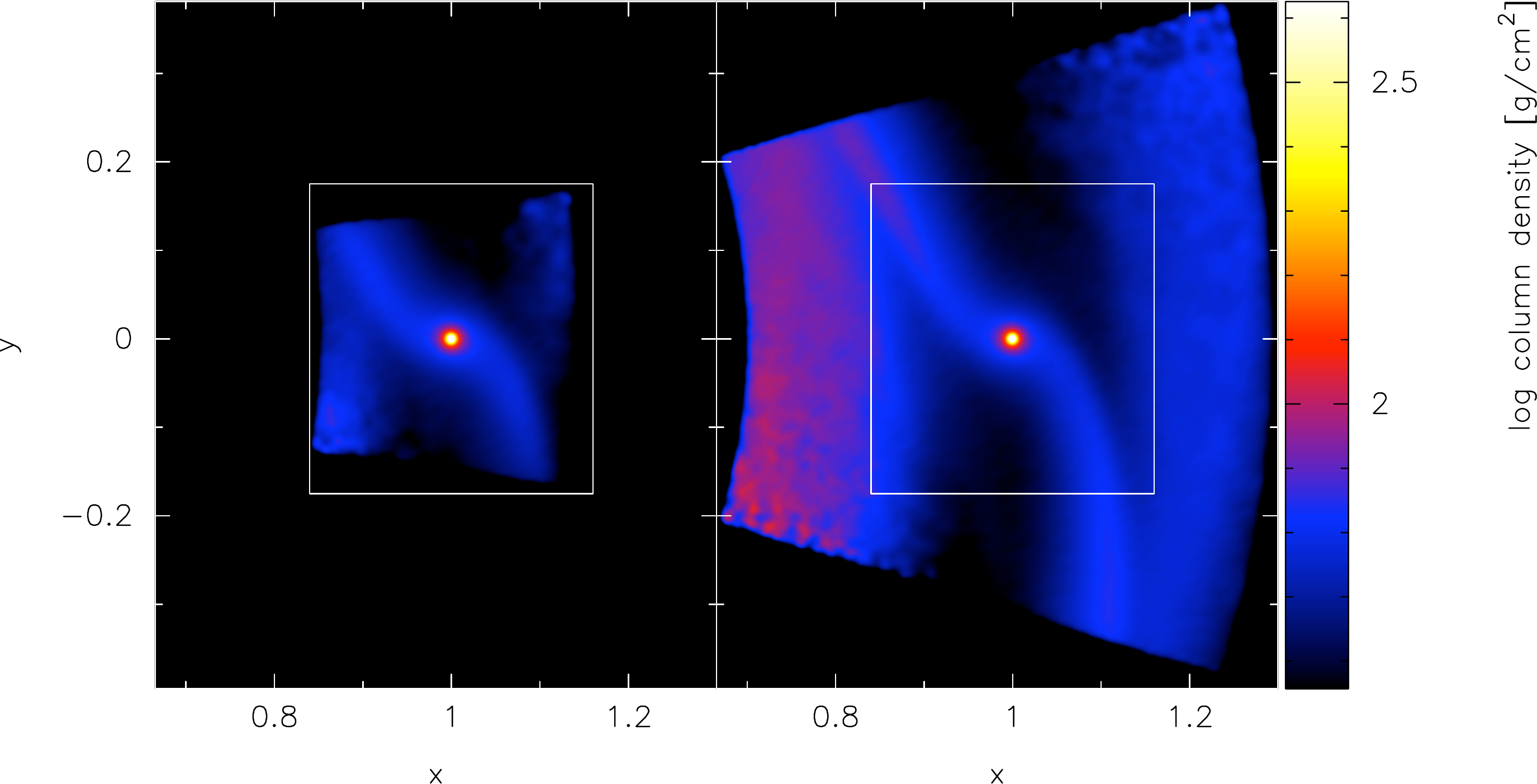}
\caption{The left panel shows a surface density plot of a standard section size calculation with a 33~$\rm M_{\earth}$ protoplanet, whilst the right panel shows an otherwise identical calculation with both the radial and azimuthal ranges doubled in extent. The calculations were performed with self-gravitating radiation hydrodynamics using standard interstellar grain opacities.  Our results do not change significantly if the disc section is enlarged. }
\label{fig:bigsec}
\end{figure*}

\begin{figure*}
\centering

\subfigure 
{
	\includegraphics[width=8cm]{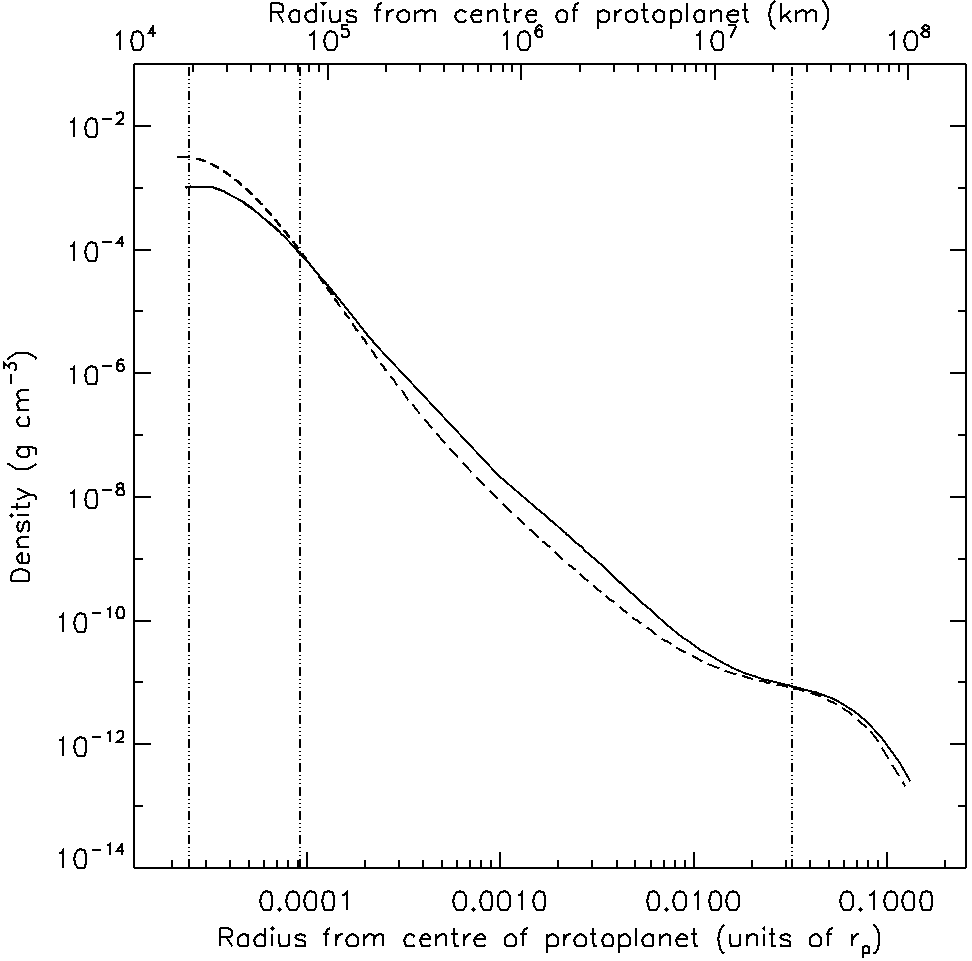}
}
\hfill
\subfigure 
{
	\includegraphics[width=8cm]{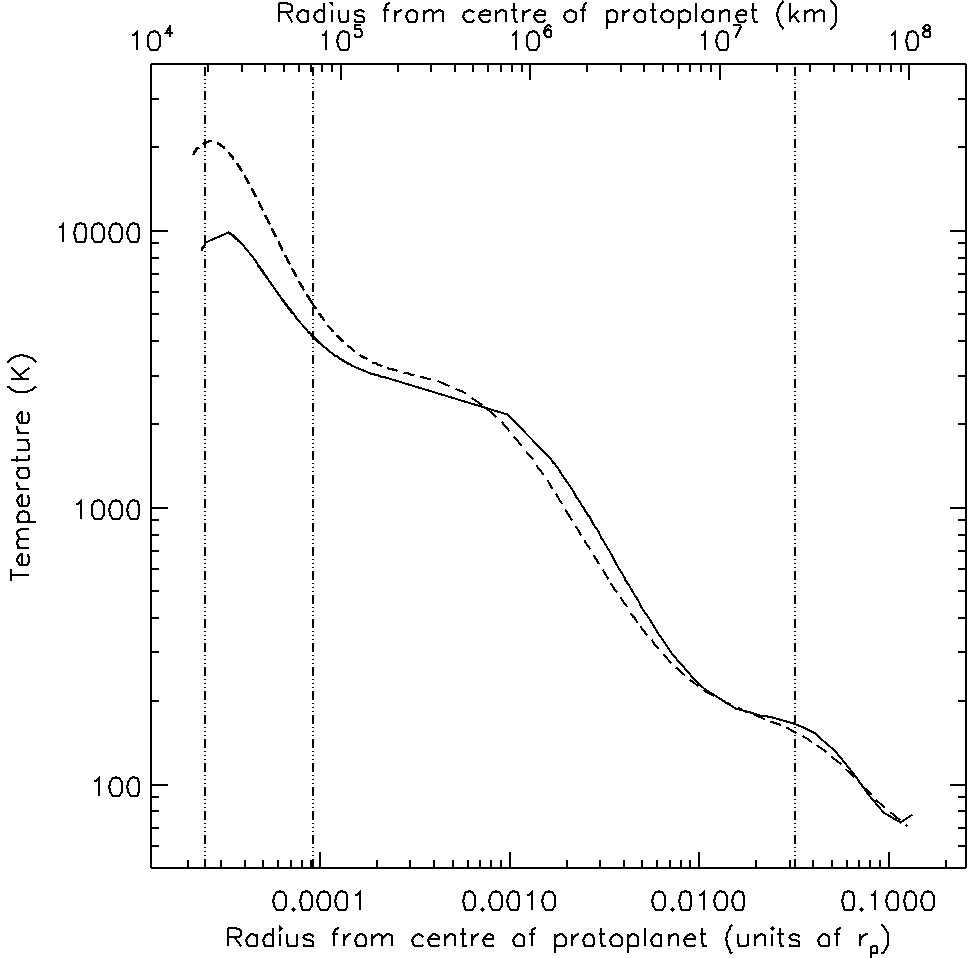}
}

\caption{The left panel shows the density fall off along the vertical ($z$) direction from the centre of a 33~M$_{\earth}$ planetary core, whilst the right panel shows the temperature fall off over the same range. The solid line in both cases represents a $3 \times 10^{4}$ particle calculation, whilst the dashed line is for a $10^{5}$ particle calculation. The calculations are both self-gravitating radiation hydrodynamical calculations using standard interstellar grain opacities and our standard protoplanetary disc. The vertical lines (dash-dots-dash) in both cases from left to right indicate the realistic solid core radius, a Jupiter radius, and the Hill radius of the core.}
\label{fig:denscomp}
\end{figure*}

\section{Tests}

In this section, we investigate the effects of changing our numerical model parameters on the results. Specifically, we compare the results from two calculations that have different disk section sizes, and we compare the results of two calculations that employ different numbers of SPH particles to model the gas.  These calculations were performed using our standard planetary core of 33~$\rm M_{\earth}$ embedded in a protoplanetary disc with our standard surface density.  

\subsection{Dependence on section size}
\label{sec:secsize}

We performed a calculation with 4 times the area of our standard section size (i.e., $r=1 \pm 0.3~r_{\rm p}$ and $\phi=\pm 0.3$ radians), with the same number density of particles within the region modelled by both calculations. Surface density plots are shown in Figure \ref{fig:bigsec} for our standard section size (left panel) and the larger section (right panel). Both calculations were using self-gravitating radiation hydrodynamics with protoplanet radii of 1\% of the Hill radius and standard interstellar grain opacities.  At 10 orbits the two models had accretion rates that differed by less than 10\%, and the masses contained within the self-consistently calculated Hill spheres differ by less than 3\%.

\subsection{Dependence on numerical resolution}
\label{numdep}

Most calculations in this paper use $3 \times 10^4$ SPH particles to model the accretion in our standard disc section. We tested the effects of resolution by running $10^{5}$ particle calculations for 33~$\rm M_{\earth}$ cores at each of the opacities that we make use of in our standard $3 \times 10^{4}$ particle calculations. For the high resolution runs and their standard resolution equivalents we use the realistic physical core radii from \cite{seager07}. The particle number density we obtained using $3 \times 10^4$ particles when scaled to the volume of the 2-20 AU disc model of \cite{FouMay2008} is equivalent to using $4 \times 10^{7}$ particles. Their maximum resolution was $10^{6}$ particles for this model. Smoothing lengths near the planetary cores in our models, which give the effective resolution length, are as low as $3 \times 10^{-4} R_{H} \approx R_{p}$, which is nearly two orders of magnitude better than the resolution of \cite{DAnLub2008}.

Figure \ref{fig:denscomp} shows the radial density and temperature profiles of the protoplanets for the two different resolutions using standard interstellar grain opacities. The structure of the protoplanets is similar for the two resolutions.  The greatest differences occur at the surface of the core with the higher resolution calculation resolving higher densities and temperatures.  The maximum density is $\approx$ 3 times greater in the higher resolution calculation, and the temperature $\approx$ 2 times higher. The standard and high resolution calculations using reduced opacities also display good agreement with each other.  We also investigated the dependence of the accretion rates on the resolution and found that the accretion rates differed by between 2\% and 10\% for all opacities with the accretion rates being calculated at 2 orbits. These calculations could not be followed for many orbits due to the computation time required to evolve these small core calculations, particularly at high resolutions where each orbit took $\sim$ 16 weeks of CPU time.

\begin{table*}
\centering
\begin{tabular}{c c c c c c c}
\hline\hline
Mass ($\rm M_{\earth}$) & Locally-isothermal  & Standard opacity & 10\% opacity & 1\% opacity & 0.1\% opacity & Standard opacity high-mass disc \\ [0.5ex]
\hline
10 & $4.6\times10^{-5}$ & $6.5\times10^{-7}$ &  $1.0\times10^{-6}$ &  $3.2\times10^{-6}$ & $4.8\times10^{-7}$ & -- \\
22 & $8.1\times10^{-5}$ & $1.1\times10^{-6}$ &  $3.6\times10^{-6}$ &  $1.0\times10^{-5}$ &  $1.4\times10^{-5}$ & $7.3\times 10^{-6}$\\
33 & $8.5\times10^{-5}$ & $2.1\times10^{-6}$ &  $6.1\times10^{-6}$ &  $2.4\times10^{-5}$ &  $4.7\times10^{-5}$ & $6.6\times 10^{-6}$\\
56 & $9.1\times10^{-5}$ & $5.0\times10^{-6}$ &  $1.7\times10^{-5}$ &  $6.0\times10^{-5}$ &  $5.7\times10^{-5}$ & --\\
100 & $9.2\times10^{-5}$ & $3.3\times10^{-5}$ &  $4.6\times10^{-5}$ &  $5.4\times10^{-5}$ &  $5.1\times10^{-5}$ & $3.3\times 10^{-5}$\\
166 & -- & $3.8\times10^{-5}$ &  $4.7\times10^{-5}$ &  $5.2\times10^{-5}$ &  $4.3\times10^{-5}$ & $1.7\times 10^{-4}$\\
333 & $8.0\times10^{-5}$ & $2.6\times10^{-5}$ &  $3.2\times10^{-5}$ &  $3.4\times10^{-5}$ &  $3.0\times10^{-5}$ & $1.3\times 10^{-4}$\\ [1ex]
\hline
\end{tabular}
\caption{The accretion rates of various mass protoplanets (in $\rm M_{\rm J}~yr^{-1}$) obtained from our SPH calculations.  Rates are given for the locally-isothermal calculations and the self-gravitating radiation hydrodynamical calculations using core radii of 1\% of the protoplanet's Hill radius with standard interstellar grain opacities and with 10\%, 1\%, and 0.1\% of these values.  The last column also gives radiation hydrodynamical accretion rates with standard interstellar grain opacities, but using a protoplanetary disc that is ten times more massive (750 g~cm$^{-2}$ at the protoplanet's radius, rather than our standard surface density of 75 g~cm$^{-2}$).}
\label{table:isorates}
\end{table*}

\section{Results}
\label{results}

\subsection{Locally-isothermal calculations}

We performed locally-isothermal calculations to provide us with upper limits for the rates of accretion on to planetary cores, and to allow us to compare our results with those of \cite{batezeus}.  With a locally-isothermal equation of state there is no heating of the gas near the planet due to the accretion. There is therefore no increase in thermal pressure to oppose the collapse of the gas onto the planetary core, and gravity is the dominant force in determining the dynamics of the system.

\subsubsection{Accretion rates}
\label{sec:iso}

\begin{figure}
\centering
\includegraphics[width=8cm]{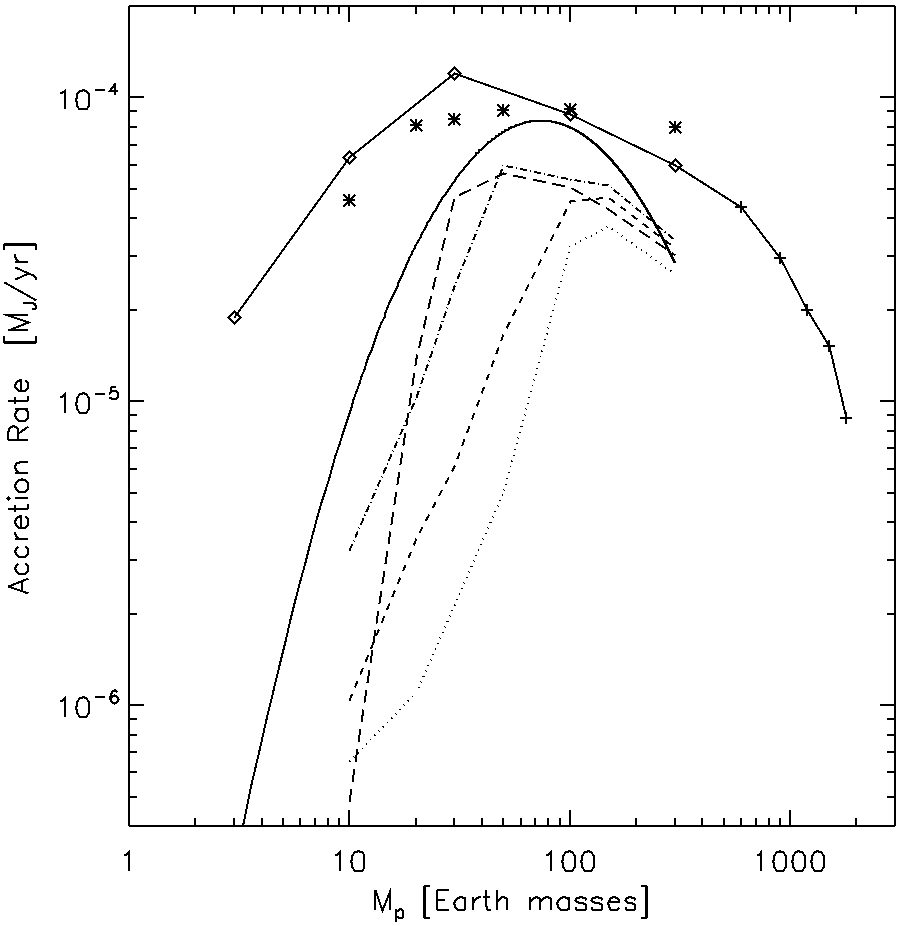}
\caption{Protoplanet accretion rates. The asterisks mark the accretion rates for our SPH calculations using a locally-isothermal equation of state, essentially providing upper limits to the accretion rates obtainable with various protoplanet masses. The diamonds mark the results of \protect \cite{batezeus}, and the plus signs those of \protect \cite{LubSeiArt1999}, connected with solid lines. These calculations were also locally-isothermal but were global disc simulations performed using the ZEUS code.  The SPH and ZEUS accretion rates are in reasonable agreement.  The accretion rates from our self-gravitating radiation hydrodynamical SPH calculations using core radii of 1\% of the protoplanet's Hill radius are given using line types that denote the different grain opacities.  Results are shown using standard interstellar grain opacities (IGO) (dotted), 10\% IGO (short-dashed), 1\% IGO (dot-dash), and 0.1\% IGO (long dashed).  The inclusion of radiative transfer substantially lowers the accretion rates of low-mass protoplanets, but Jupiter-mass protoplanets have similar accretion rates to the locally-isothermal result, regardless of the grain opacity. The analytic approximation of \protect \cite{angelo2003} is shown by the solid curved line.
}
\label{fig:accrate}
\end{figure}

Table \ref{table:isorates} lists the accretion rates obtained using the locally-isothermal equation of state for 6 different core masses and our standard protoplanetary disc. These accretion rates are also plotted with asterisks in Figure \ref{fig:accrate} (amongst other data). The locally-isothermal accretion rates increase with the core mass below $\sim 20$~M$_{\earth}$, level off above this mass, and eventually begin decreasing above $\sim 100$~M$_{\earth}$. This turn over with increasing core mass is a result of the disc gap that broadens for higher mass cores, acting to limit the flow of gas into the Hill sphere. A turn over in accretion rate with increasing core mass was also obtained by \cite{batezeus} using ZEUS, and their results, as well as those of \cite{LubSeiArt1999}, are plotted in Figure \ref{fig:accrate} for comparison (the solid lines connecting diamond and plus symbols). The SPH and ZEUS accretion rates always lie within 30\% of each other, although the turn over from the SPH calculations is not as abrupt as was obtained in the ZEUS calculations.  This level of agreement is satisfactory given that the main aim of this paper is to determine how the inclusion of radiative transfer affects the accretion rates and, as will be seen, radiative transfer can alter the accretion rates by up to two orders of magnitude!  The reason for the differences between the ZEUS and SPH calculations is not clear, but it is probably due to a combination of the different viscosities in the two types of calculation and the fact that although the ZEUS calculations have settled to a quasi-equilibrium protoplanetary disc structure, injecting SPH particles into the disc section based on this ZEUS quasi-equilibrium will not produce an SPH disc structure that is exactly in quasi-equilibrium. The smooth solid curve plotted in Figure \ref{fig:accrate} is the analytic approximation of \cite{angelo2003} based on their locally isothermal three-dimensional calculations, and represents upper limits for accretion using their model.

Isothermal calculations were also performed that used a surface density of 750$\rm g~cm^{-2}$ at 5.2 AU, ten times the value in our standard surface density. As expected, accretion rates under these conditions increased linearly with the disc surface density.

Our isothermal results give upper limits for the accretion rates that we can expect to find. Radiative transfer (and other effects such as heating of the envelope due to planetesimal bombardment) is only likely to decrease the accretion rates.

\begin{figure*}
\centering
\subfigure 
{
    \includegraphics[width=7cm]{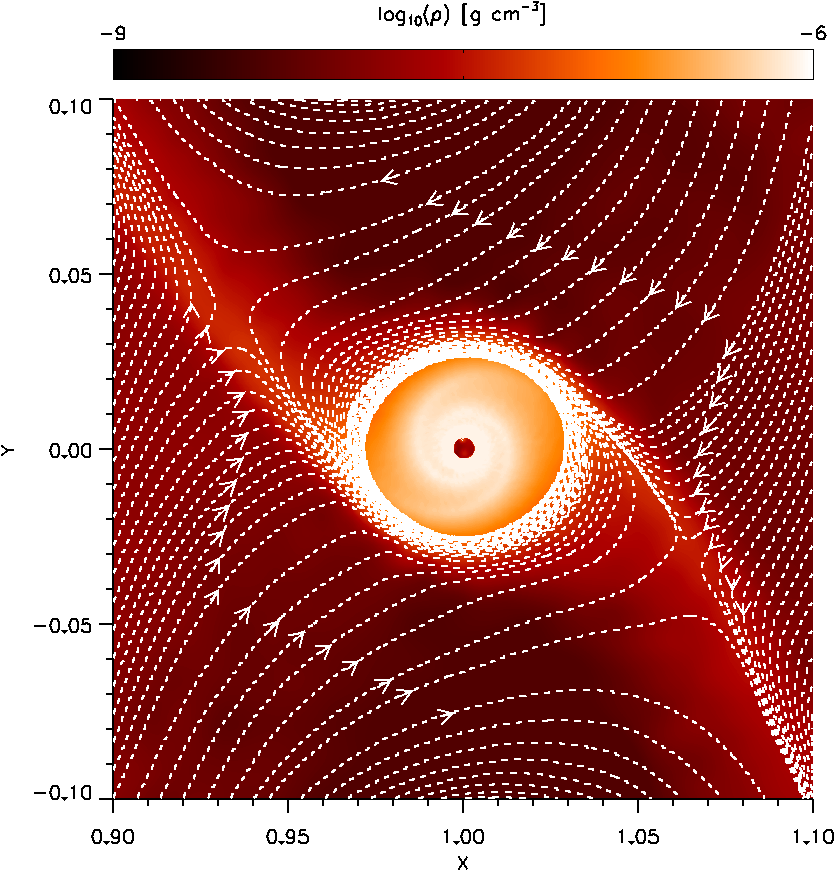}
}
\subfigure 
{
    \includegraphics[width=7cm]{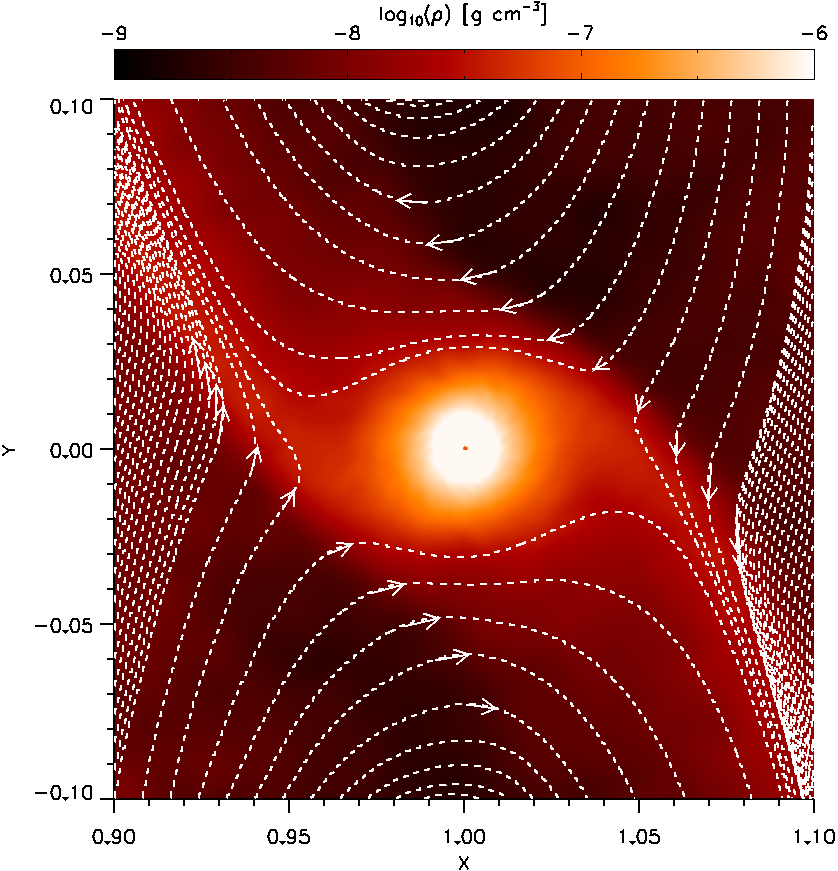}
\vspace{-6pt}
}
\subfigure 
{
    \includegraphics[width=7cm]{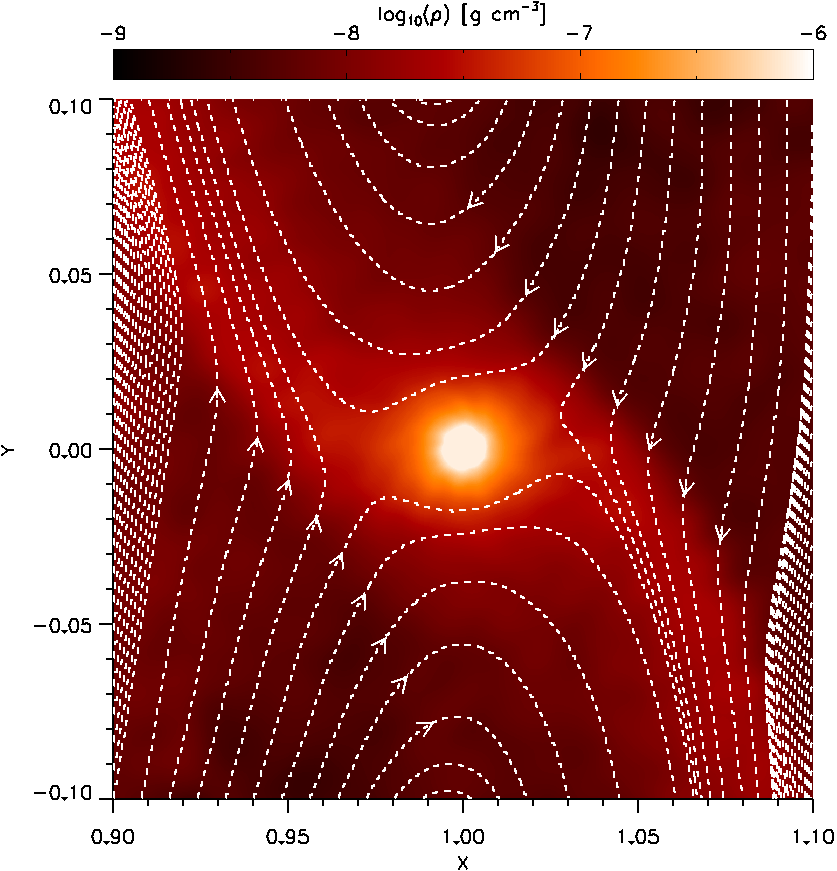}
}
\subfigure 
{
    \includegraphics[width=7cm]{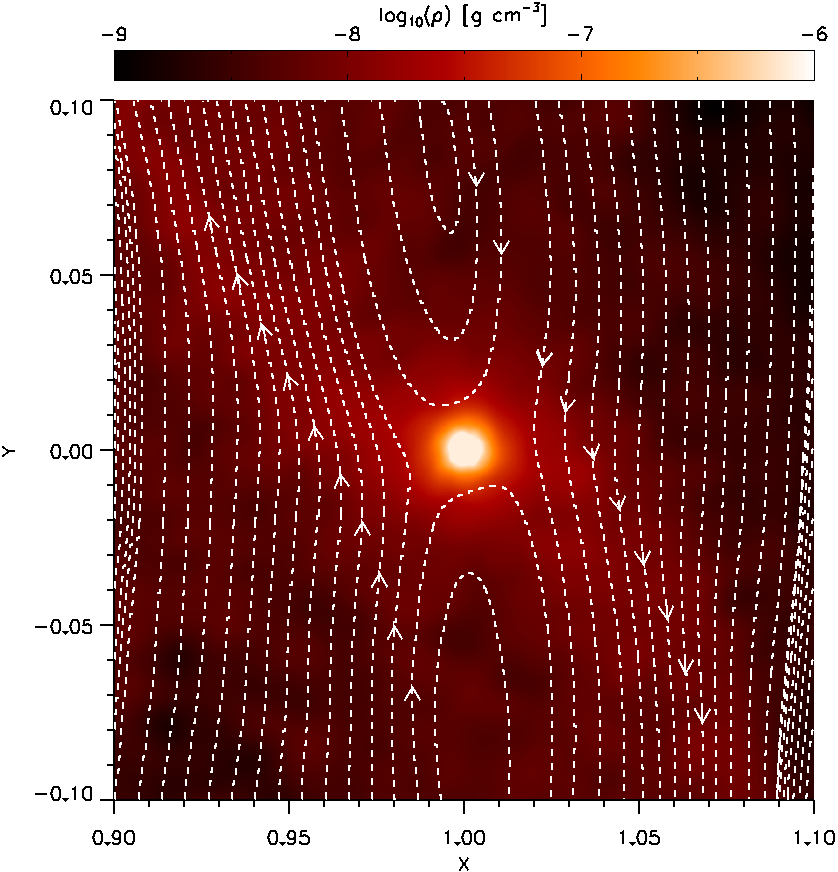}
}

\caption{Disc density ($\rho$) and streamlines at the disc midplane for a range of protoplanet masses: 333, 100, 33, and 10 $\rm M_{\earth}$ from top-left to bottom-right.  These plots are similar to those presented in Figure 5 of \protect \cite{batezeus}.  The calculations use a locally-isothermal equation of state.  The spiral shocks and horseshoe orbit regions generated by the protoplanets are clearly visible.  Circumplanetary discs are formed in all but the lowest mass case (this is readily apparent in the highest mass case from the streamlines).}
\label{fig:streams}
\end{figure*}

\subsubsection{Gas dynamics}

Along with the accretion rates, we compare the gas dynamics in the SPH calculations with those \cite{batezeus} obtained using ZEUS.  Figure \ref{fig:streams} plots the midplane density and gas streamlines for the 10, 33, 100, and 333 M$_{\earth}$ cores in a similar manner to Figure 5 of \citeauthor{batezeus}.
The spiral shocks generated by the protoplanet in the SPH simulations closely resemble those of \citeauthor{batezeus}, with sharp changes in gas velocities at the shock fronts, and gas spiralling on to the core via circumplanetary discs.  \citeauthor{batezeus} resolved circumplanetary discs in their locally-isothermal calculations for protoplanets with masses as low as 33~M$_{\earth}$, but they did not have the resolution to investigate whether discs formed in their lower mass cases.  For our SPH calculations, the presence of circumplanetary discs is clearly demonstrated by the steamlines in the Jupiter-mass case, but smaller discs are also present for the lower-mass protoplanets, again down to 33~M$_{\earth}$.

\begin{figure*}
\centering
\subfigure 
{
    \includegraphics[width=7cm]{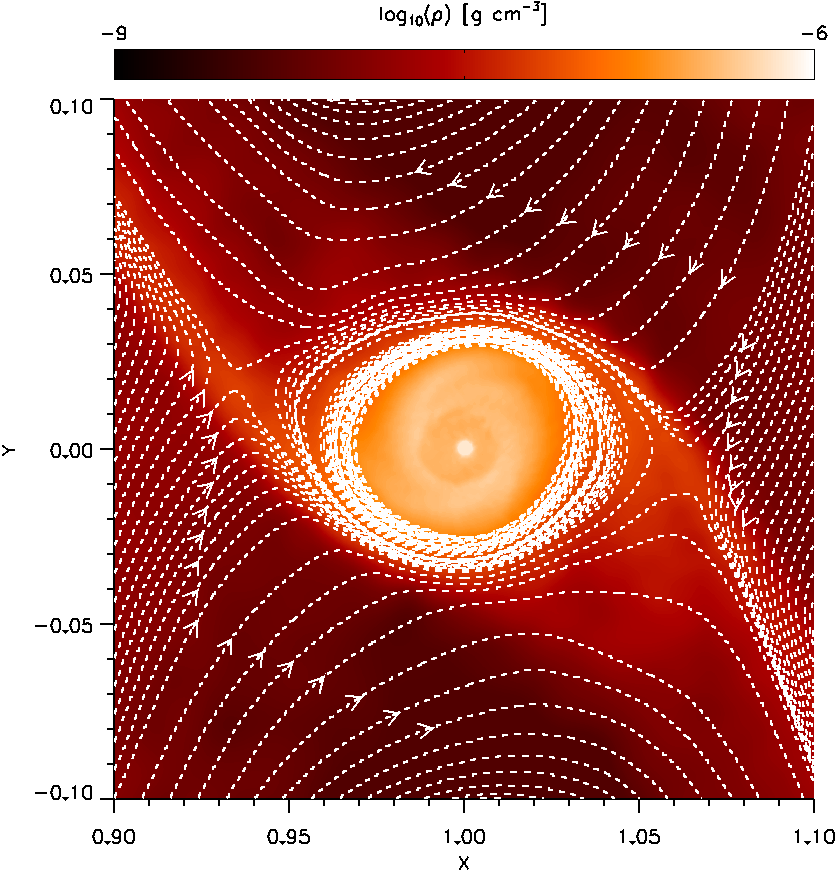}
}
\subfigure 
{
    \includegraphics[width=7cm]{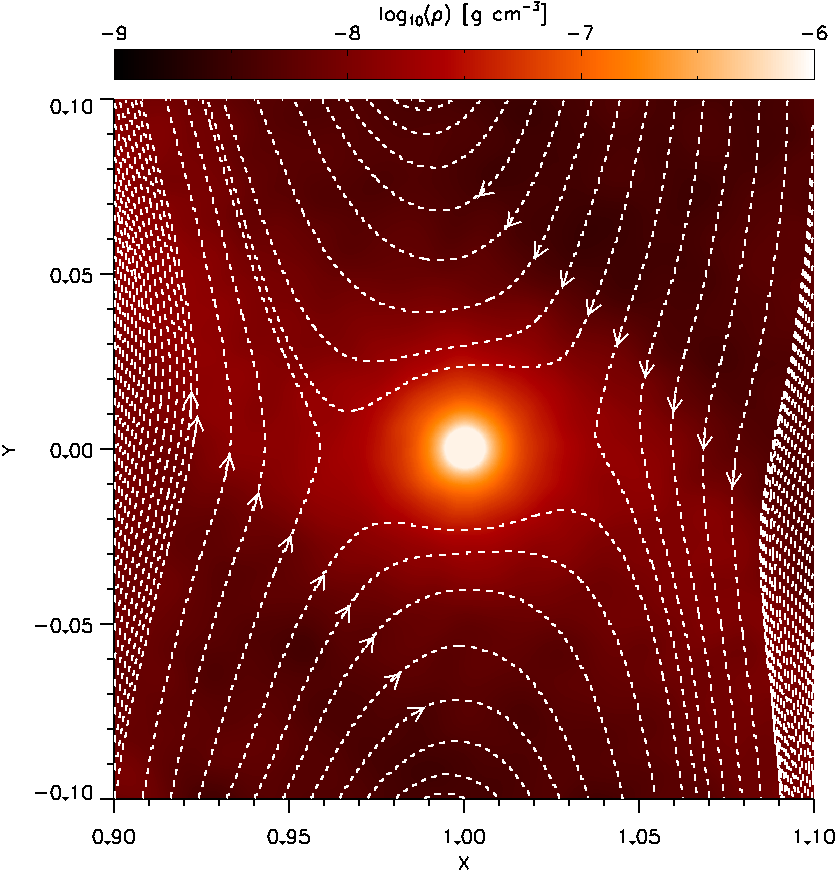}
}

\subfigure 
{
    \includegraphics[width=7cm]{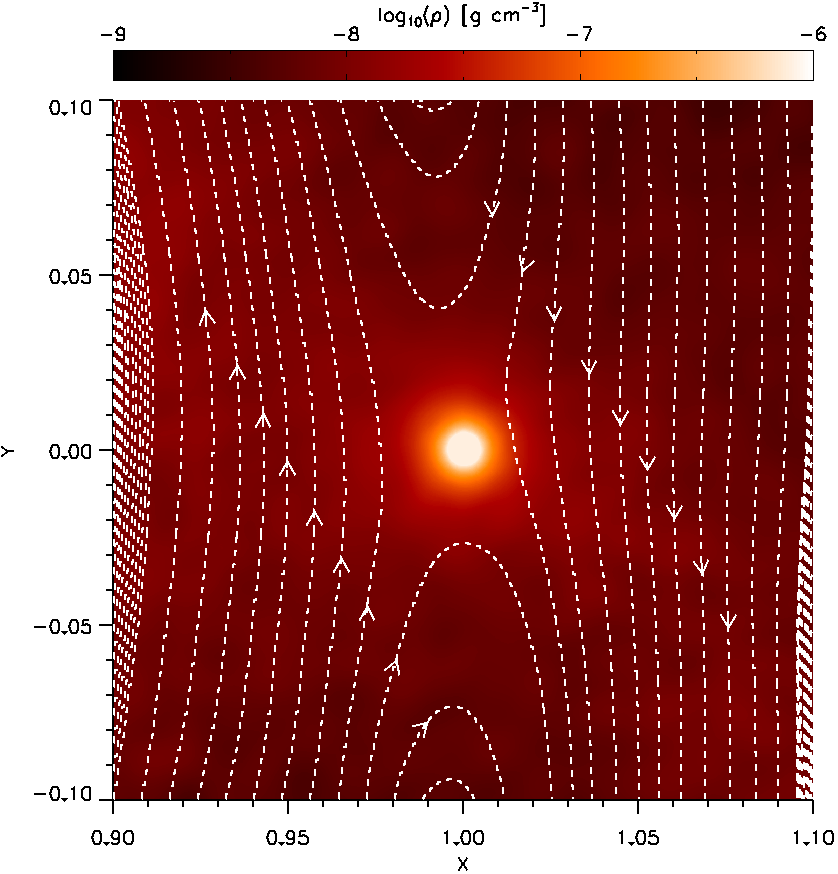}
}
\subfigure 
{
    \includegraphics[width=7cm]{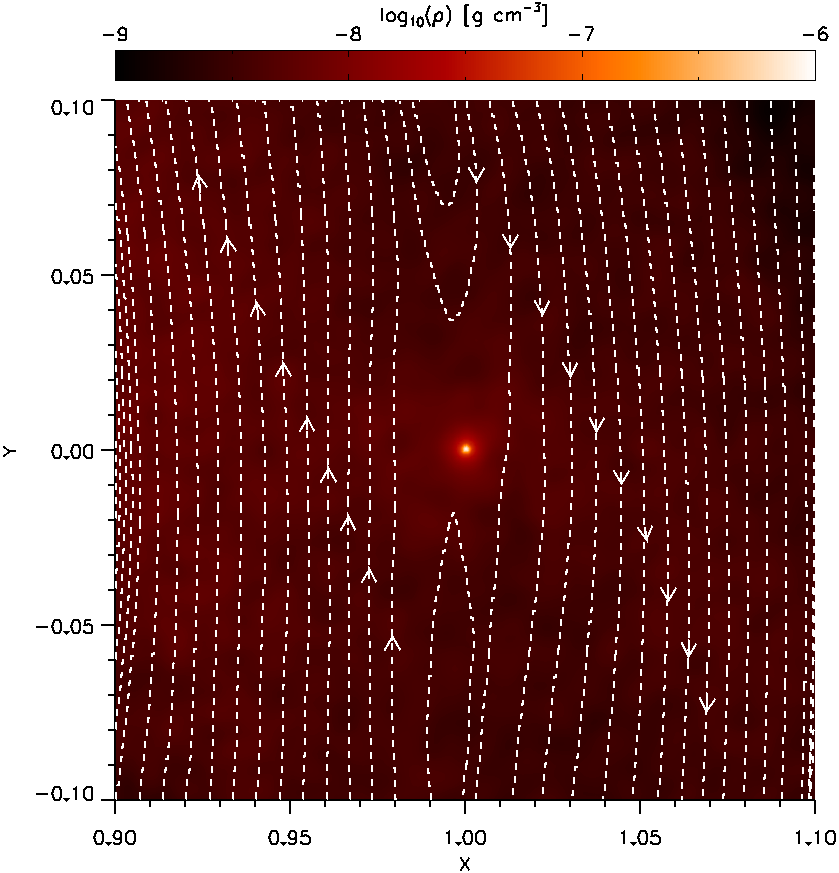}
}

\caption{Disc density ($\rho$) and streamlines at the disc midplane for the self-gravitating radiation hydrodynamical calculations using core radii of 1\% of the Hill radii and standard interstellar grain opacities for the same protoplanet masses as in Figure \ref{fig:streams}.  The highest mass case is similar to that obtained using the locally-isothermal equation of state, but the circumplanetary disc is somewhat larger and the spiral shocks within it are less pronounced.  For the lower mass protoplanets, however, the spiral shocks in the protoplanetary disc are much weaker and the horseshoe orbit regions are much narrower in the radiation hydrodynamical calculations than in the locally-isothermal calculations.}
\label{fig:radstreams}
\end{figure*}

\begin{figure*}
\centering
\subfigure 
{
    \includegraphics[width=7cm]{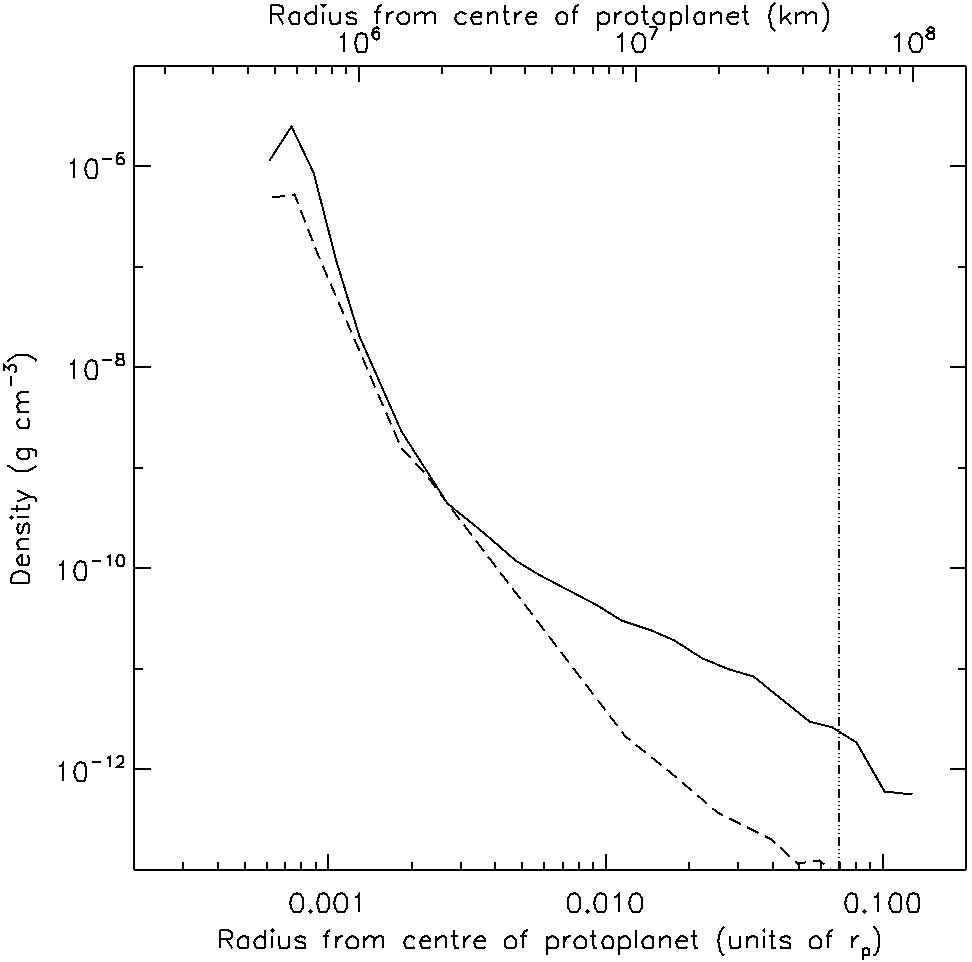}
}
\subfigure 
{
    \includegraphics[width=7cm]{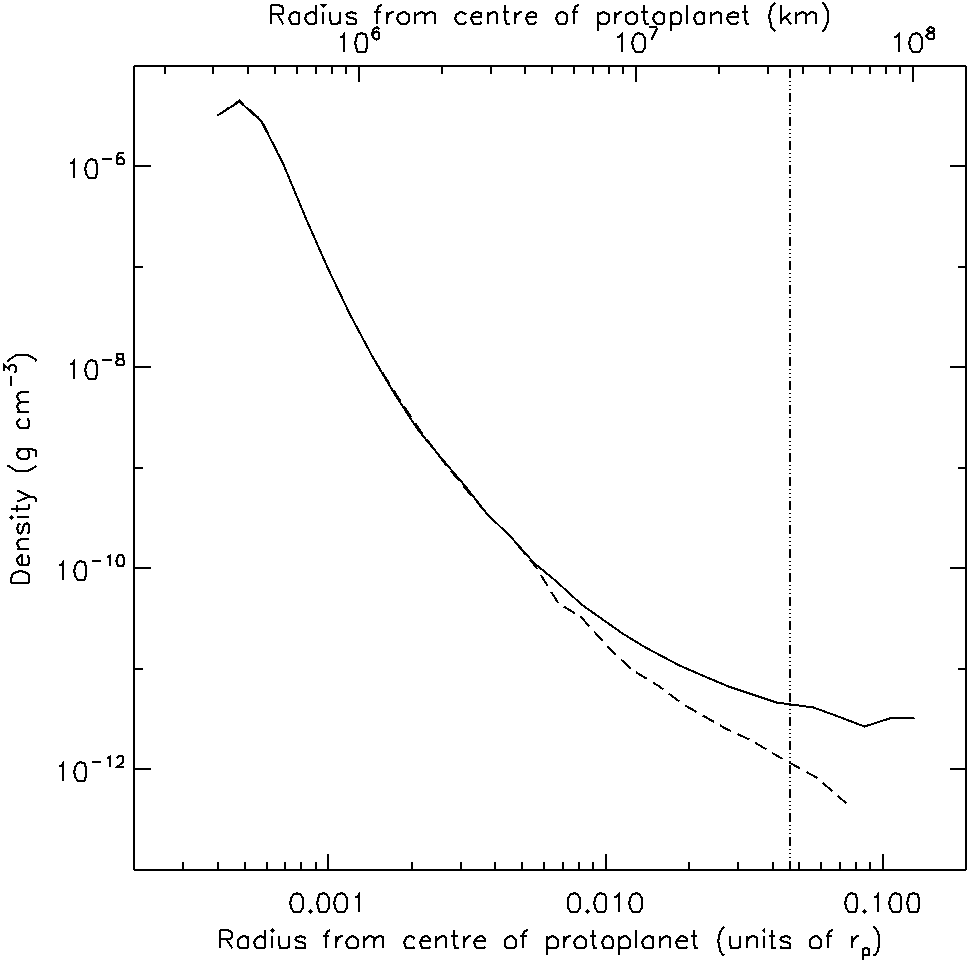}
}

\subfigure 
{
    \includegraphics[width=7cm]{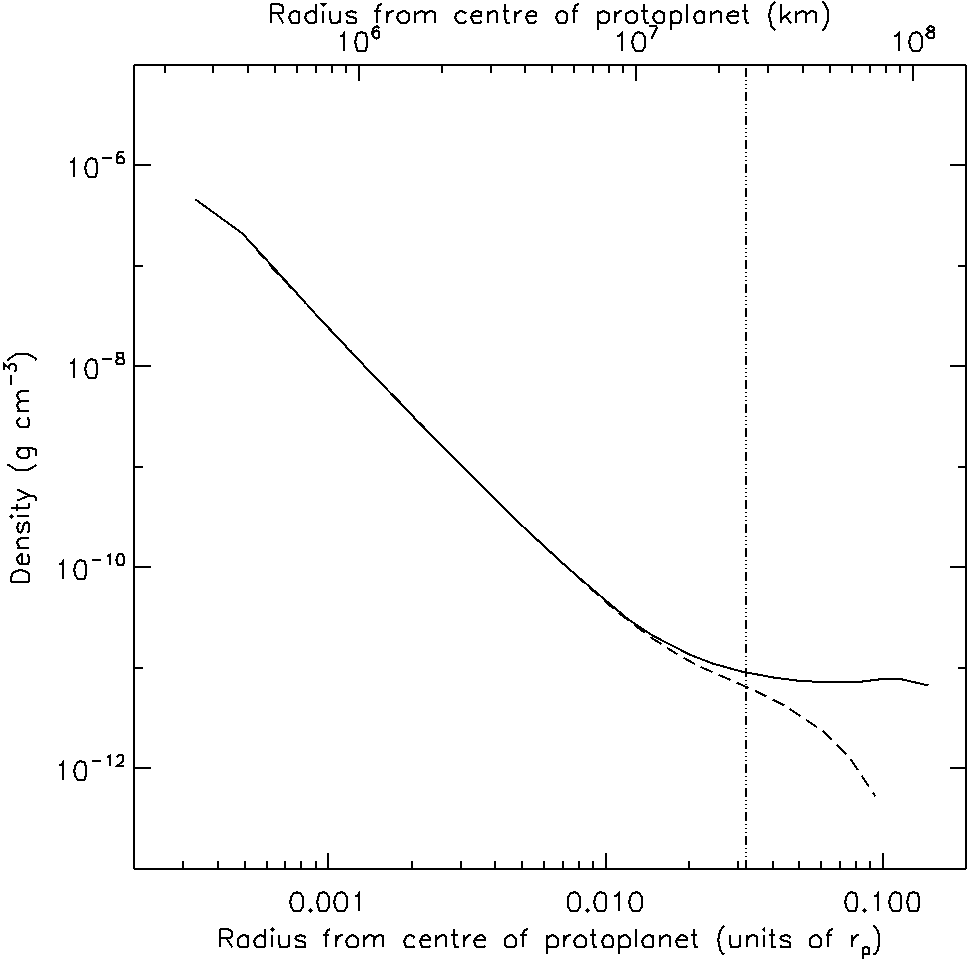}
}
\subfigure 
{
    \includegraphics[width=7cm]{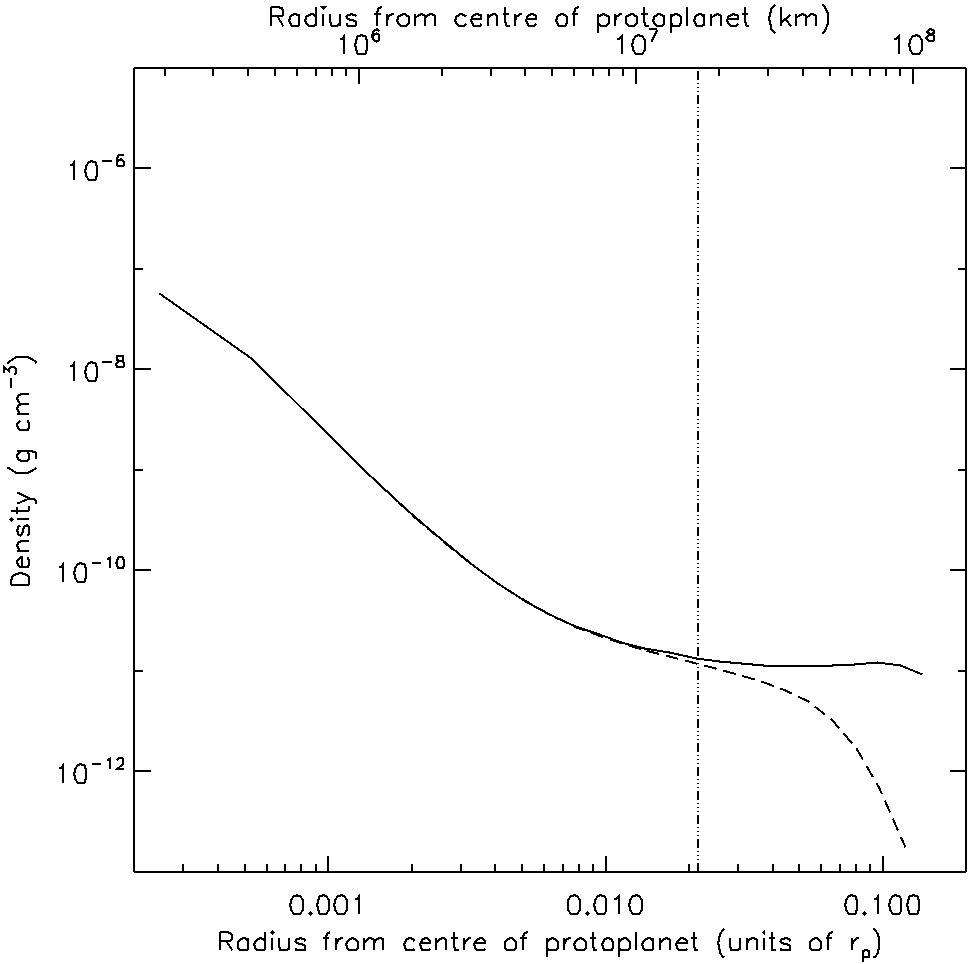}
}

\caption{Density ($\rho$) profiles of the accreting protoplanets for the self-gravitating radiation hydrodynamical calculations using core radii of 1\% of the Hill radii and standard interstellar grain opacities for the same protoplanet masses (333, 100, 33, and 10~M$_{\earth}$) as in Figures \ref{fig:streams} and \ref{fig:radstreams}.   The solid lines give the density profiles along the $x$-axis in the disc midplane from the planetary core towards the central star.  The dashed lines give the vertical density profiles (along the $z$ direction).  It is clear from these density profiles that the 333~M$_{\earth}$ protoplanet forms a circumplanetary disc, and the 100~M$_{\earth}$ protoplanet has a flattened envelope within their respective Hill radii (marked by the vertical dash-dots-dash lines). The lower mass protoplanets are surrounded by almost spherical envelopes.}
\label{fig:densityprofiles}
\end{figure*}

\subsection{Self-gravitating radiation hydrodynamical calculations}

In locally-isothermal calculations, thermal pressure plays a very minor role in determining the gas dynamics near the protoplanet.  In reality, however, the release of gravitational energy during the accretion process heats the gas near the protoplanet which lends thermal support to the gas and inhibits accretion. This process can only be treated properly by including radiative transfer.  The main question we investigate in this paper is, how much is the accretion rate decreased from the locally-isothermal rate when the process is modelled using self-gravitating radiation hydrodynamics?

\subsubsection{Gas dynamics}
\label{sec:gasdynamics}

\begin{figure*}
\centering

\subfigure 
{

\setlength\fboxsep{0pt}
\setlength\fboxrule{0.0pt}
\fbox{\includegraphics[width=14cm]{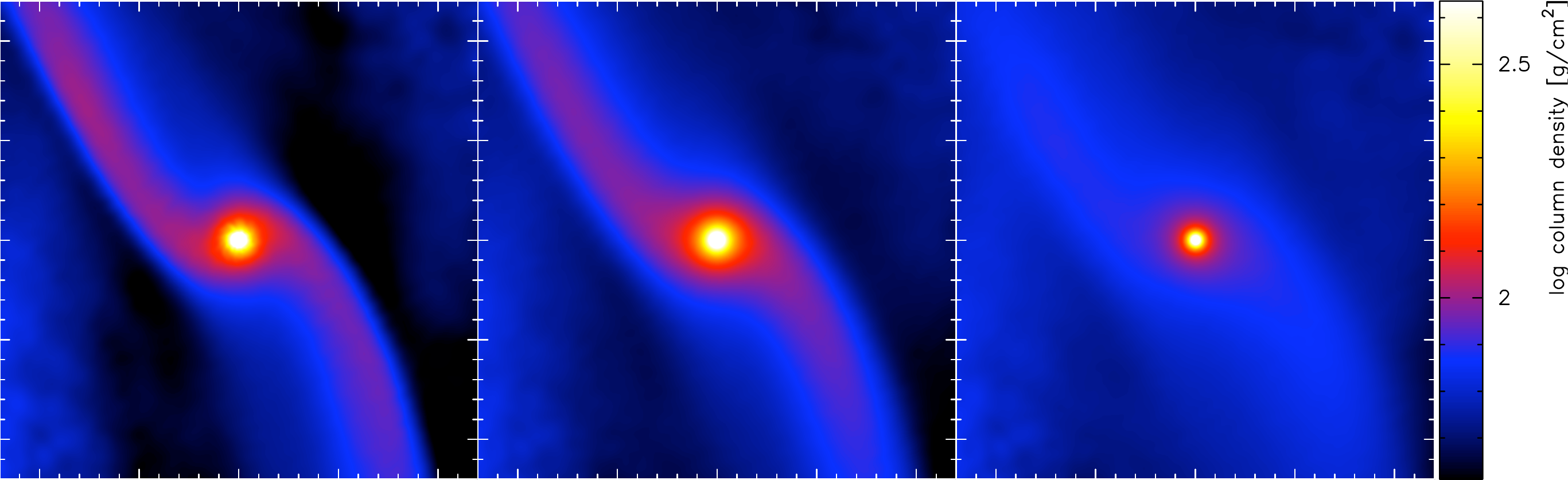}}
}

\vspace{-4mm}

\subfigure 
{

\setlength\fboxsep{0pt}
\setlength\fboxrule{0.0pt}
\fbox{\includegraphics[width=14cm]{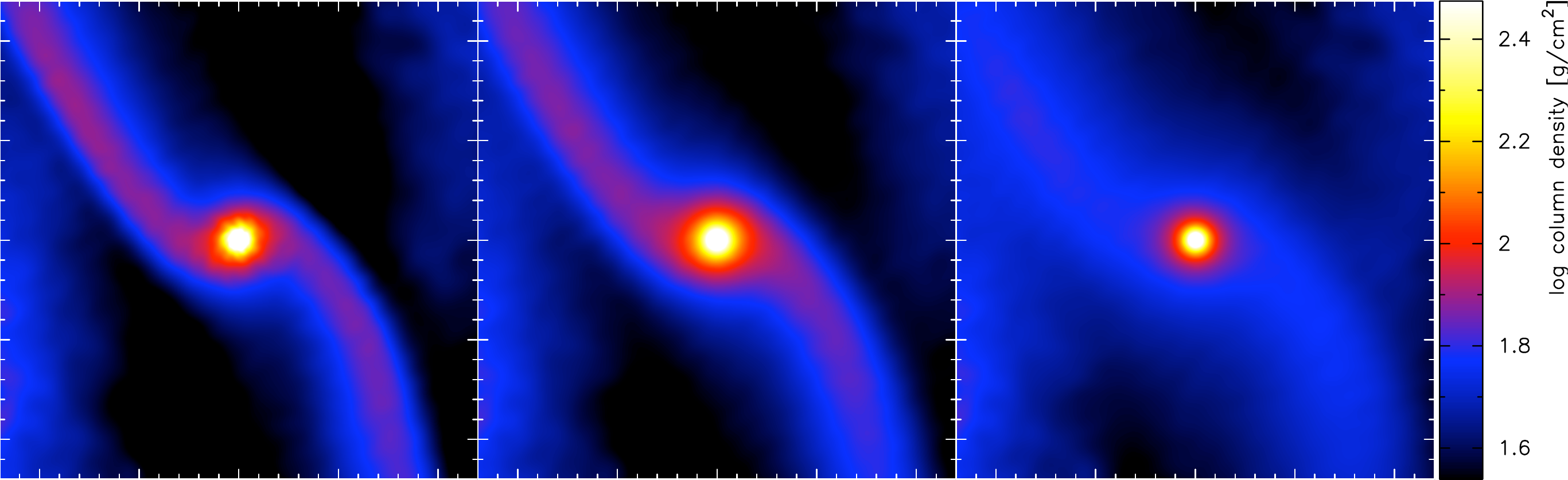}}
}

\vspace{-4mm}
\subfigure 
{
        \setlength\fboxsep{0pt}
\setlength\fboxrule{0.0pt}
\fbox{\includegraphics[width=14cm]{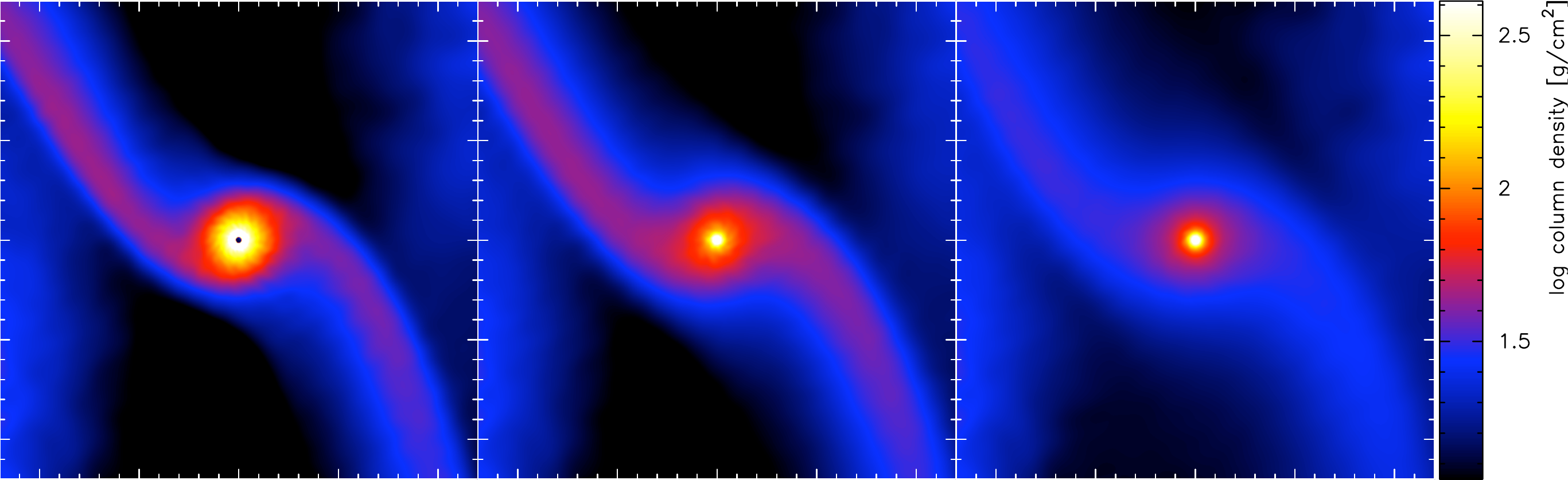}}
}

\vspace{-4mm}
\subfigure 
{
        \setlength\fboxsep{0pt}
\setlength\fboxrule{0.0pt}
\fbox{\includegraphics[width=14cm]{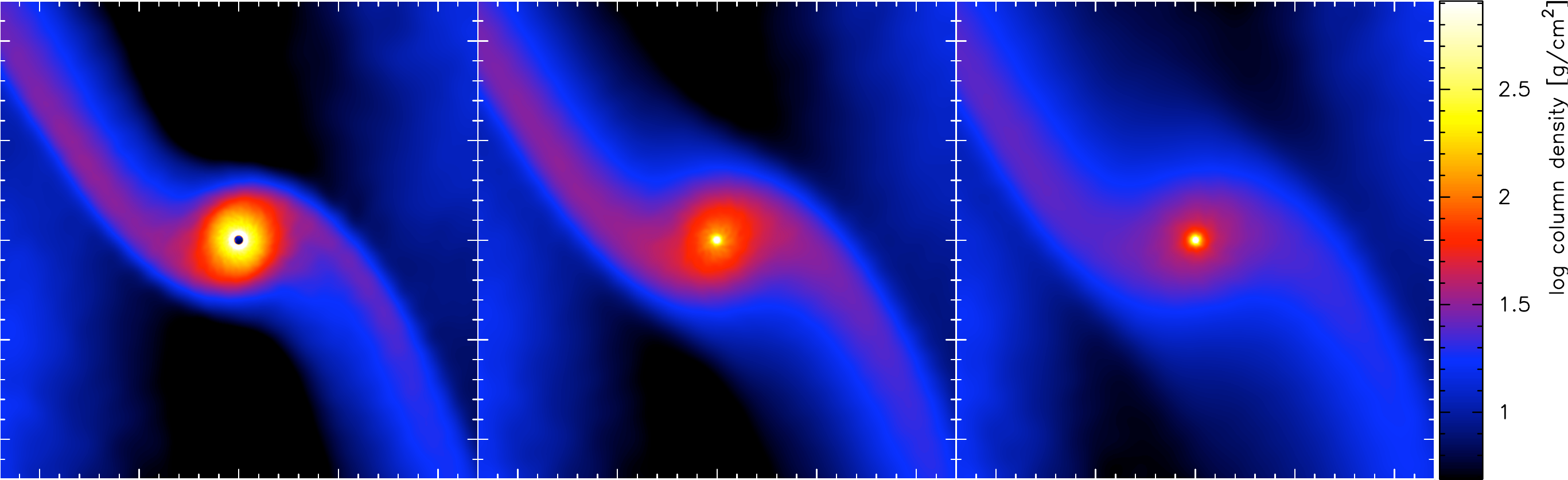}}
}

\vspace{-4mm}

\subfigure 
{
            \setlength\fboxsep{0pt}
\setlength\fboxrule{0.0pt}
\fbox{\includegraphics[width=14cm]{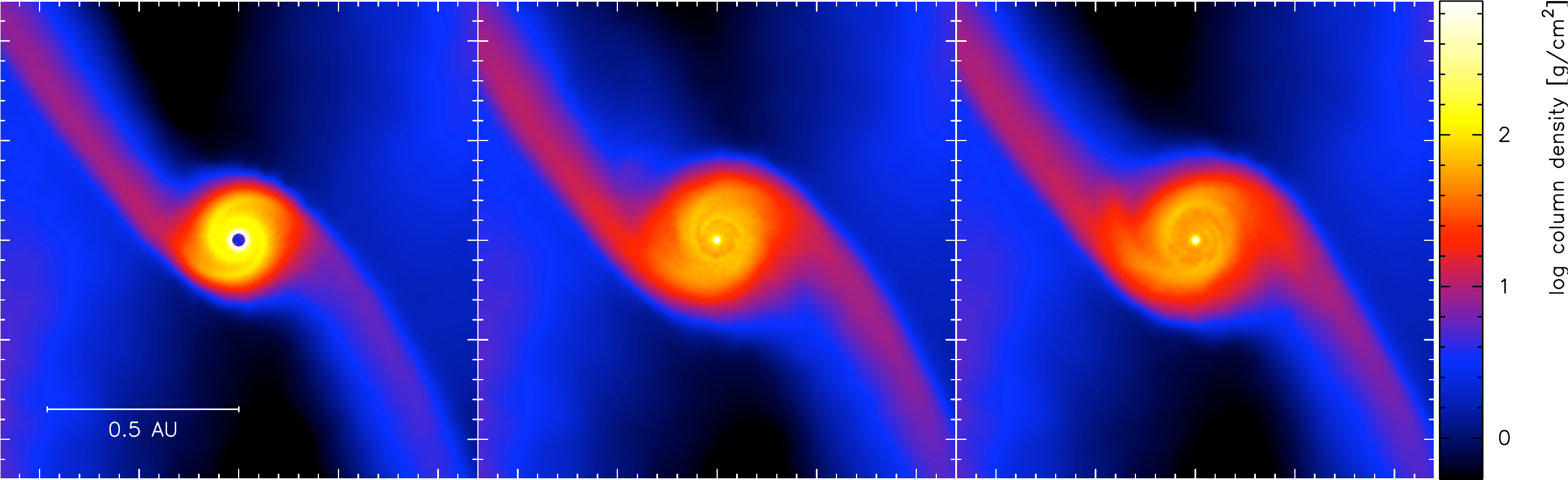}}
}

\caption{Surface density plots for locally-isothermal calculations (left), and self-gravitating radiation hydrodynamical calculations using 1\% IGO (centre), and standard IGO (right) calculations with our standard protoplanetary disc surface density. From top to bottom the protoplanet masses are 22, 33, 100, 166 and 333 $\rm M_{\earth}$ respectively.  The radiation hydrodynamical calculations use protoplanet radii of 1\% of the Hill radii, while the locally-isothermal calculations use 5\% of the Hill radii.  Note that with radiative transfer and standard opacities, the spiral shocks in the protoplanetary disc are much weaker than using a locally-isothermal equation of state, while using radiative transfer with reduced grain opacities results in intermediate solutions because the discs are less optically thick and are able to radiate more effectively than with standard opacities.}
\label{fig:surfdens}
\end{figure*}

Figure \ref{fig:radstreams} shows density plots and streamlines for self-gravitating radiation hydrodynamical calculations of the same four protoplanet masses as those shown in Figure \ref{fig:streams}.  These calculations all have full IGO and our standard protoplanetary disc.  The Jupiter-mass protoplanet has very similar gas dynamics in its vicinity regardless of the thermal treatment. This is because the gravitational forces are dominant over the thermal pressures involved. The only significant changes are that the circumplanetary disc is somewhat hotter, thicker, and larger in radius and the spiral shocks that are prominent in the locally-isothermal circumplanetary disc are much weaker in the radiation hydrodynamical calculation (see also Figure \ref{fig:surfdens}). At lower core masses, where the gravitational forces are weaker, the increased thermal pressure in the vicinity of the protoplanet smears out the features seen under locally-isothermal conditions. The spiral shocks become much less prominent and the radial extents of the horseshoe orbit regions are greatly decreased. There is no perceptible deflection of the streamlines near the 10~$\rm M_{\earth}$ protoplanet, while although the shocks and their associated streamline deflections are visible in the 33 and 100~$\rm M_{\earth}$ protoplanet cases, they are much weaker than in the locally isothermal case. Finally, the mass distribution within the Hill radius of protoplanets with masses $\lsim 100$~M$_{\earth}$ is found to be an envelope rather than a circumplanetary disc.  Whereas the 166 and 333~M$_{\earth}$ cases have a clear disc, the lower-mass protoplanets are surrounded by an envelope with very little vertical flattening (Figure \ref{fig:densityprofiles}).

With a fixed planet orbital radius and only modelling a small section of the protoplanetary disc we are unable to investigate the torque exerted on the protoplanet by the disc and its consequent radial migration rate.  However, for low-mass cores ($\rm < 100M_{\earth}$) the substantial weakening of the shocks with the inclusion of radiative transfer suggests that their migration rates should be substantially reduced in comparison with locally-isothermal migration models. Indeed, \cite{MorTan2003}, \cite{PaaMel2006}, \cite{PaaMel2008}, and \cite{kley2008} all find that the radial migration of protoplanets are altered significantly when non-isothermal calculations are performed reducing the inward radial migration rate and in some cases producing outward radial migration.

\subsubsection{The effect of reduced grain opacities}

In the previous sections, we presented results from locally-isothermal calculations and radiation hydrodynamical calculations using full interstellar grain opacities (IGO).  In this section, we investigate intermediate cases.  It is expected that as the opacity of the gas is decreased, the radiation hydrodynamical calculations should become more like the locally-isothermal calculations.

In Figure \ref{fig:surfdens}, we provide surface density plots for locally-isothermal calculations (left) and radiation hydrodynamical calculations with 1\% IGO (centre) and full IGO (right) for core masses of 22, 33, 100, 166, and 333 M$_{\earth}$.  As expected, there is a clear transition from strong isothermal shocks to weak shock fronts as the opacity increases and radiative transfer becomes more important. Also, we find that a circumplanetary disc is formed around a 100 M$_{\earth}$ protoplanet when the opacity is reduced to 1\% IGO whereas when the full IGO is used the protoplanet is surrounded by a flattened envelope.  This reflects the increased ability of the lower opacity gas to cool.  However, even using 1\% IGO, 56 M$_{\earth}$ and lower mass protoplanets are unable to form discs and are surrounded by envelopes.

As mentioned above, although we are unable to calculate the radial migration rates of the protoplanets from these calculations, we nevertheless predict that the radial migration rates in radiation hydrodynamical simulations will depend on the grain opacity of the gas because of the opacity dependence of the spiral shock structure.  Lower grain opacities will lead to inward migration rates more similar to the locally-isothermal migration rates, while high opacity gas will lead to slower inward migration rates.  This is a further complication in trying to understand the already complex problem of protoplanet migration in a protoplanetary disc.

\subsubsection{Accretion rates}
\label{funcform}

In the midplane of the protoplanetary disc, gas accreted by the protoplanet comes from the region between the horsehoe orbit streamlines and the region where streamlines are perturbed by the protoplanet, but nevertheless continue to orbit the star (Figures \ref{fig:streams} and \ref{fig:radstreams}).  The widths of these regions decrease as the protoplanet's mass decreases and, thus, the protoplanet accretion rate also decreases.  For higher-mass protoplanets the resulting accretion rate is not as large as would be predicted purely by the width of the accreting streamline region because the accretion rate also depends on the gas density in this region which, for high-mass protoplanets, is dramatically decreased by the opening of the gap in the disc by the protoplanet.  Thus, for very high-mass protoplanets ($\gsim 333$~M$_{\earth}$) the accretion rate quickly decreases with increasing mass (Figure \ref{fig:accrate}).

As mentioned in Section \ref{sec:gasdynamics}, the inclusion of radiative transfer decreases the size of the horseshoe orbit region for the low-mass protoplanets and, simultaneously, the width of any accreting streamline region.  This leads directly to greatly reduced accretion rates for the low-mass protoplanets compared to the locally-isothermal calculations.  Another way of understanding the effect is that the additional thermal support provided by the gravitational energy released by the accreting gas inhibits further gas accretion onto the protoplanet with radiative transfer. Only as thermal energy is radiated out of the envelope can more gas be accumulated by the protoplanet.

The degree of reduction of the accretion rates below the locally-isothermal upper limits depends upon the opacity of the gas, and the mass of the protoplanet. Figure \ref{fig:accrate} illustrates the accretion rates obtained for 7 core masses (10, 22, 33, 56, 100, 166, and 333$\rm M_{\earth}$) using self-gravitating radiation hydrodynamics for 4 different grain opacities (full, 10\%,1\%, and 0.1\% interstellar grain opacities) and our standard protoplanetary disc.

At the low mass end (excepting the 0.1\% IGO, $\rm 10 M_{\earth}$ case), where thermal effects rival the importance of gravity, the accretion rates increase with decreasing opacity. The accretion rate is dependent upon the condensation of gas towards the core, which is in turn limited by the rate at which energy can be removed by radiation. Reducing the opacity reduces the optical depth to the protoplanet allowing energy to be released more rapidly and the gas to cool more quickly, which in turn leads to faster accretion by the protoplanet. As the mass of the protoplanet increases so the dependence of the accretion rate on the opacity decreases, evidence of gravity's increasingly dominant influence. At 333~$\rm M_{\earth}$ the accretion rates at all the different opacities are very similar and are all approaching the isothermal rate, where gravity is the dominant force at work.

The accretion rate for a $\rm 10~M_{\earth}$ protoplanet in an envelope with a 1/1000 IGO gives the slowest accretion rate that we find. In this case, the grain opacity has become so low that radiative emission from the dust grains is inefficient in cooling the gas. The gas outside of the protoplanet's envelope is essentially isothermal with a temperature greater than that in locally-isothermal models or the other radiation hydrodynamical models with larger grain opacities.  This hotter gas is less readily captured by the protoplanet and so the accretion rate is particularly low.

In Table \ref{table:isorates}, we also list the accretion rates obtained from radiation hydrodynamical calculations using protoplanetary discs that are an order of magnitude more massive (750~g~cm$^{-2}$) than our standard case.  These calculations used standard IGO.  For locally-isothermal calculations (Section \ref{sec:iso}), the accretion rates scaled linearly with the disc's surface density.  However, we see from Table \ref{table:isorates} that for the radiation hydrodynamical calculations, increasing the surface density by an order of magnitude increases the accretion rate by much less than an order of magnitude.  The greater gas densities make it more difficult for the envelope to radiate away its thermal energy and accept more gas so that the accretion rates do not increase linearly with the disc surface density. 

\begin{figure}
\centering
\includegraphics[width=8cm]{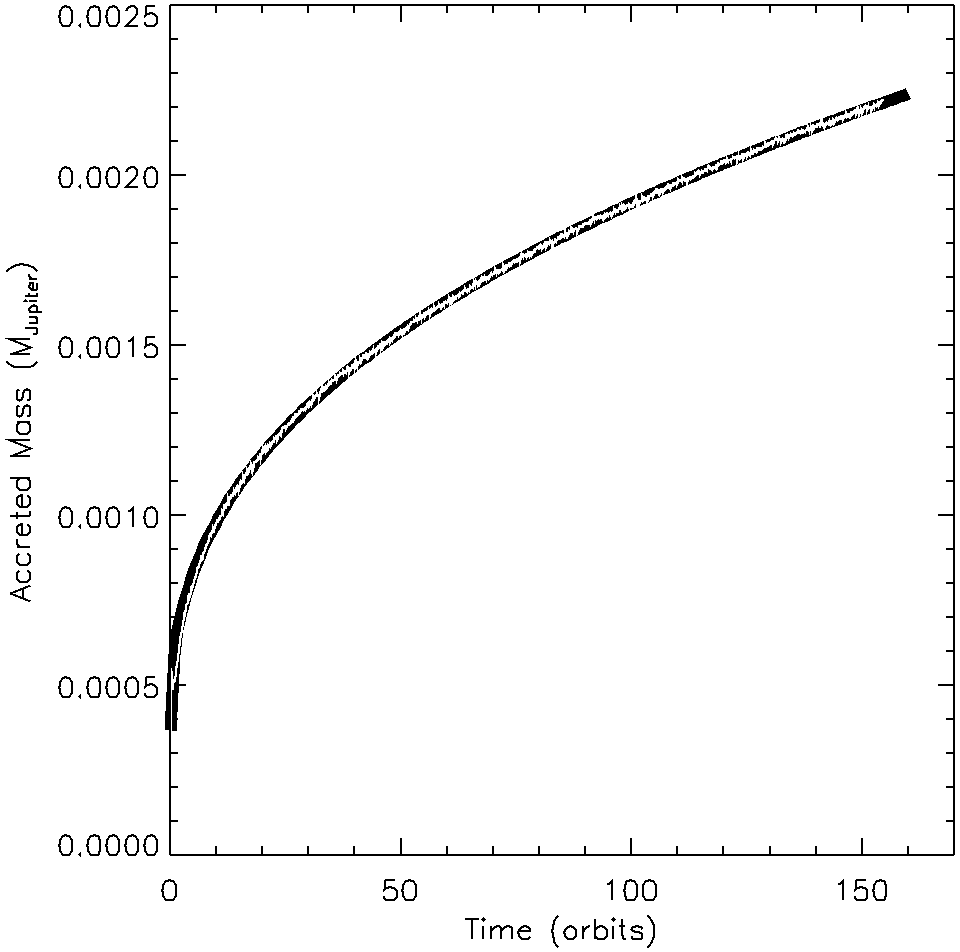}
\caption{The white line shows the accreted gas mass measured over the course of $\sim$160 orbits for a 33~$\rm M_{\earth}$ core embedded in our standard protoplanetary disc with standard interstellar grain opacities, and with a core radius of 1\% $\rm R_{H}$. The thick black line over which the white line can be seen is a fit with the functional form $M=at^{b}+c$ where $b=0.40$.}
\label{fig:longacc}
\end{figure}

The accretion rates discussed here are measured at ten orbits. Although they have reached a quasi-equilibrium state by then, they all display a slow decrease with time, i.e. $\rm \ddot M_{\rm H} < 0$. The accretion rates slowly decline because the envelope finds it increasingly difficult to cool as it grows in mass (due to the increasing optical depth).  To investigate this slow decline we managed to evolve our standard case to 160 orbits (which took $\sim$ 36 CPU weeks). Figure \ref{fig:longacc} illustrates that the accretion rate continues to decline with time, $t$. The curve of Figure \ref{fig:longacc} can be fitted well with the functional form $M_{\rm H}=at^{b}+c$. Using the Levenberg-Marquardt method to minimise chi-squared (omitting the first 10 orbits to avoid the influence of initial transience) we find that the curve is best fit with an exponent of $b=0.40$. Thus, the accretion rate is given by $\dot M_{\rm H} \propto t^{-0.6}$. Similar fitting can be performed, though with a shorter baseline, for other calculations some of which reached 20 orbits. Of these, the higher opacity cases can be well fitted with $b=0.40$. In lower opacity calculations the accretion rates falls away less rapidly, more resembling the isothermal cases.  The locally-isothermal calculations {\em do} attain steady rates after the first orbit. In all cases, the accretion rates cannot carry on declining indefinitely.  Instead, as the core mass is augmented with more gas so gravity becomes more dominant and the fall off is expected to cease.  Then accretion rates will then start increasing, eventually leading to runaway accretion.  We discuss the long-term evolution of the accretion rates further in Section 5.

\subsubsection{Realistic resolved solid cores}

In the above sections, we have used protoplanets with radii of 1\% of their Hill radii.  These are $10-20$ times larger than the realistic radii of solid planetary cores (Table \ref{table:radii}), but they have the advantage that they require $\approx 2$ orders of magnitude less computational time!

Here we present three-dimensional self-gravitating radiation hydrodynamics models of true core accretion that resolve the flow of gas all the way down to realistically-sized planetary cores. This is the first time such calculations have been performed.  As well as being the correct way to model protoplanet accretion, the calculations allow us to explore the validity of results obtained with our 1\% Hill radius protoplanets. In these calculations, we use a 33~$\rm M_{\earth}$ core with a radius of just 3.1 $\rm R_{\earth}$ (as specified in Table \ref{table:radii}).

\begin{figure}
\centering
\includegraphics[width=8cm]{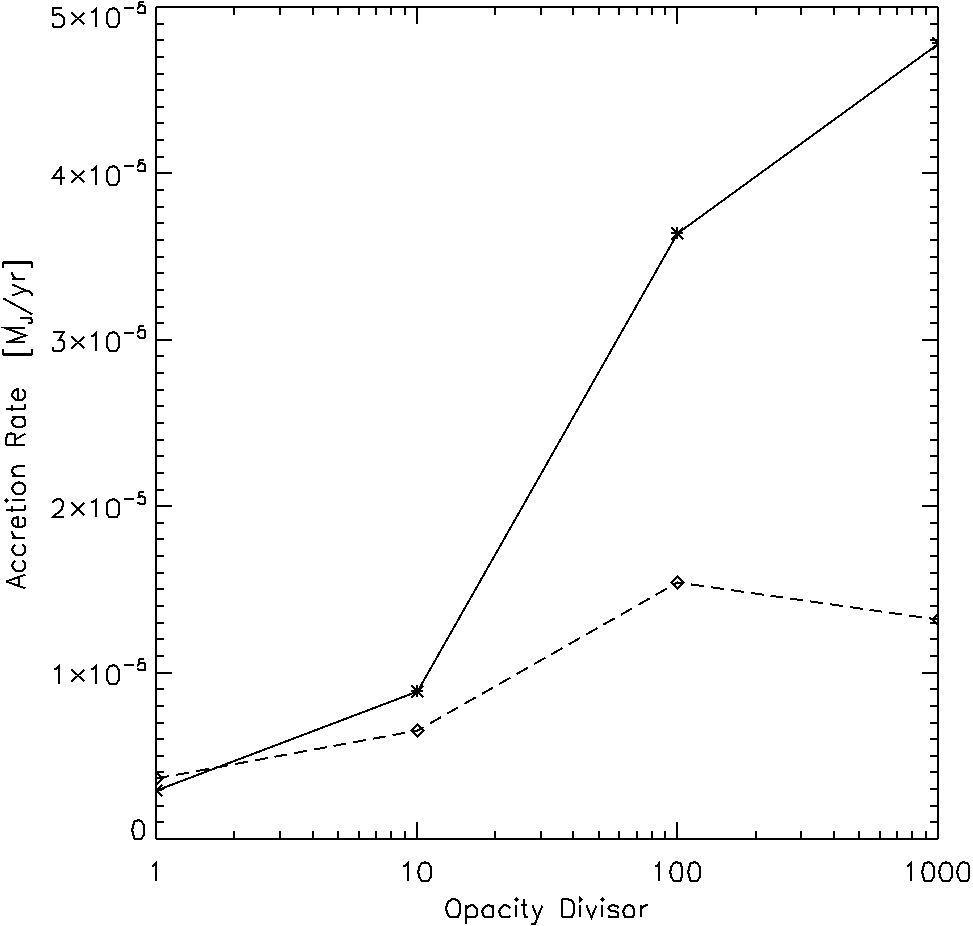}
\caption{Accretion rates for 33~$\rm M_{\earth}$ protoplanets using self-gravitating hydrodynamical calculations and four different opacities. The asterisks joined with a solid line mark the accretion rates obtained when using a protoplanet radius of 1\% of the Hill radius, whilst the diamonds connected with the dashed line show the rates obtained when using a realistic solid core radius of 3.1~R$_{\earth}$ (based on the models of \protect \citealt{seager07}).  Using a larger than realistic planetary core radius has little effect on the accretion rates obtained using standard and 10\% opacities, but when the opacities are reduced by factors of 100 and 1000, using realistic core radii results in lower accretion rates.}
\label{fig:hillvsmall}
\end{figure}

Figure \ref{fig:hillvsmall} illustrates the differences in accretion rates at 10 orbits between the different protoplanet core radii at different opacities. At standard opacity, the rate of accretion onto the two different sized protoplanets is very similar. At this opacity, the process of gas condensation within the envelope towards the core's surface is slow due to the difficulty in radiating away the gravitational energy released. Although gas can condense further into the gravitational potential well of the core when using a smaller core radius, the long timescale required to transport energy out from these depths under optically thick conditions means that the energy release rate deep within the envelope is low compared to the energy release rate further out in the envelope.  Thus, the region deep within the envelope does not play a large role in setting the gas temperature in the vicinity of the Hill radius where the capturing of gas by the protoplanet takes place and the accretion rate is essentially independent of the exact core radius used.

Reducing the opacity of the gas leads to larger accretion rates with both large and physical core radii.  However, the increase in accretion rates obtained with lower opacities is more significant for the 1\% Hill radius cores than the realistically-sized cores. As the opacity is reduced, the increase in accretion rates is tempered to a growing degree by the influence of gravitational energy being released by the condensing gas deep within the core's gaseous envelope.  With a less optically thick envelope the radiation emitted deep within the envelope can escape more quickly, leading to a faster release rate of gravitational energy which in turn increases the temperature of gas in the vicinity of the Hill radius.

The overall differences in accretion rates between the realistically-sized cores and the 1\% Hill radius models are small for full and 10\% IGO. However at lower opacities the effect of smaller cores and deeper gravitational potential wells become important, suggesting that the accretion rates derived using the 1\% Hill radii cases should be treated as upper limits for the 1\% and 0.1\% IGO cases with core masses $\lsim 100$~M$_{\earth}$.  Above 100~M$_{\earth}$, changing the grain opacity (or even switching to a locally-isothermal equation of state) has little effect on the gas accretion rate, so altering the size of the core is also unlikely to change the rates significantly.

\section{Discussion}
\label{discussion}

The growth time of giant planets is a vital test of formation models with measured protoplanetary disc lifetimes giving a well defined upper bound. Changing conditions such as the grain opacity of the disc and its density, or the initial mass of the solid planetary cores can have a substantial impact on the gas accretion rate and in turn on the growth time.

In estimating the accretion rates onto protoplanets, past papers have tended to use one of two methods.  Either one-dimensional quasi-static self-gravitating radiative transfer models of the protoplanets themselves have been used, with the surrounding protoplanetary disc being simply a boundary condition \citep[e.g.][]{BodPol1986,pollack96, hubickyj2005, papa2005}, or hydrodynamical simulations of protoplanets embedded in protoplanetary discs, but with grossly simplified physics (e.g. locally-isothermal equations of state) \citep[e.g.][]{LubSeiArt1999,batezeus,DAnHenKle2003}.  As we have shown in this paper, three-dimensional locally-isothermal hydrodynamical calculations give accurate accretion rates only for the most massive protoplanets ($\gsim 1$ Jupiter mass).  Below this, radiative transfer must be taken into account, and the accretion rates depend on the opacity and, particularly for low opacities, the assumed radius of the planetary core.

Recently, \cite{FouMay2008} have also performed self-gravitating radiation hydrodynamical SPH simulations of Jupiter-mass protoplanets embedded in protoplanetary discs.  They do not have a core radius as such, rather they soften the gravitational potential of the protoplanet with a length scale ranging from $0.2-1.0$ Hill radii.  Their underlying disc models are similar to ours with surface densities at the protoplanet's orbital radius of 75 or 150 g~cm$^{-2}$, but steeper radial profiles of $\Sigma \propto r^{-1.5}$.  For their locally-isothermal 75 g~cm$^{-2}$ models they obtain protoplanet accretion rates similar to ours and those obtained with ZEUS by \cite{batezeus} (averaged over $\sim 170$ orbits they obtain a rate of $5.5 \times 10^{-5}$ M$_{\rm J}$~yr$^{-1}$, a factor of $\sim 1.5$ different to our result). It should be noted that their accretion rates are strongly resolution dependent.  We consider their $2 \times 10^{5}$ particle calculations as both isothermal and radiative transfer results are given at this resolution. As mentioned in Section \ref{numdep}, our effective resolution is more than two orders of magnitude higher in terms of particle number compared with these particular models.  With radiative transfer and standard grain opacities, they find the accretion rate drops to $2.2\times 10^{-5}$ M$_{\rm J}$~yr$^{-1}$, which is less than a factor of 1.2 different from our own result.  Their decrease in accretion rate between isothermal and radiative transfer calculations for this very high mass core is a factor 2.5, whilst ours is 3.1.

\cite{PaaMel2008} have recently performed grid-based three-dimensional radiation hydrodynamical calculations of 1 and 5 M$_{\earth}$ planetary cores embedded in protoplanetary discs.  These are lower masses than we have considered, however, they find that the inclusion of radiative transfer decreases the accretion rates of these cores by more than an order of magnitude over the rates obtained from locally-isothermal calculations, in qualitative agreement with our results.

\begin{figure}
\centering
\includegraphics[width=8cm]{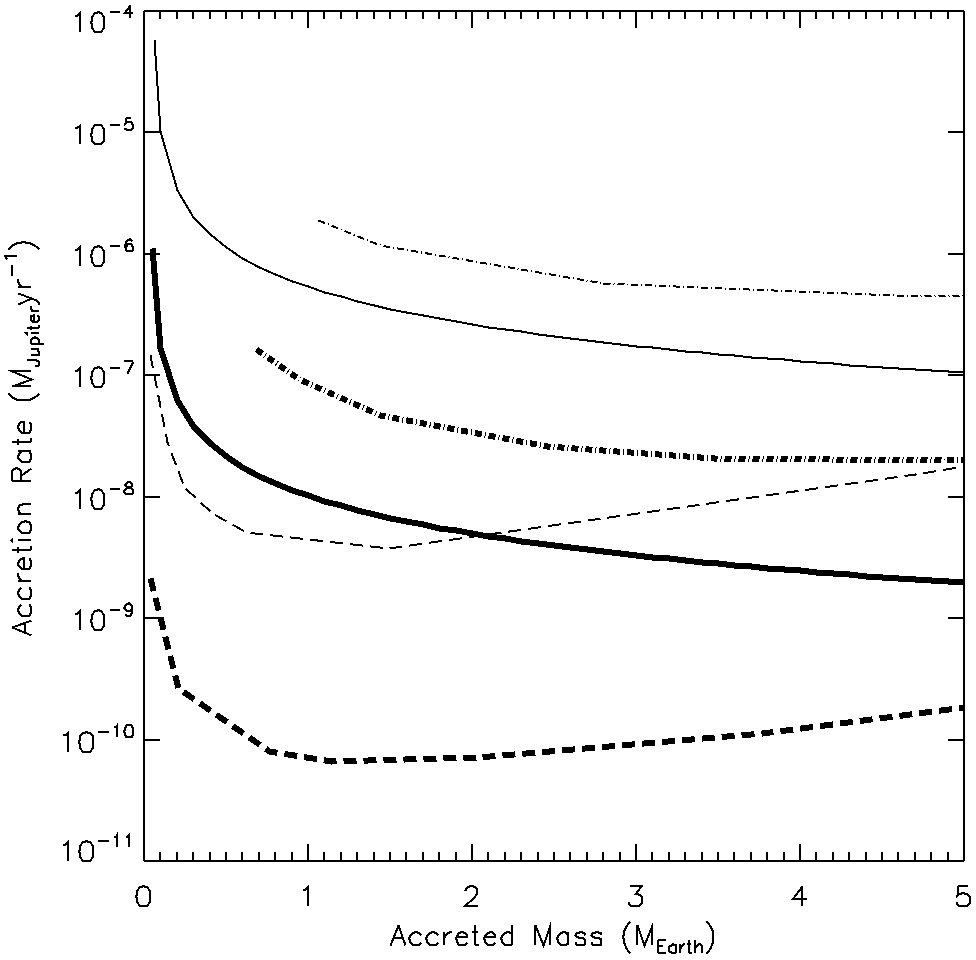}
\caption{Accretion rates versus the gas mass accreted by planetary cores for 1\% (thin lines) and standard (thick lines) grain opacities.  We plot the results of \protect \cite{papa2005} for 5~M$_{\earth}$ (dashed lines), and 15~M$_{\earth}$ (dot-dashed lines) cores. The solid lines show the extrapolations of our 10~M$_{\earth}$ core results.  Our 10~M$_{\earth}$ core accretion rates lie between the accretion rates obtained by \citeauthor{papa2005} for their 5~M$_{\earth}$ and 15~M$_{\earth}$ cores.  The effect of the reduced opacities is also consistent with \citeauthor{papa2005}'s results.}
\label{fig:papacomparison}
\end{figure}

It is somewhat more difficult to compare our results with the one-dimensional quasi-static models because we are unable to follow the growth of a protoplanet from its initial solid core mass all the way to a gas giant planet.   Furthermore, different authors begin with different solid core masses and other assumptions. Despite these difficulties, we can compare our accretion rates onto protoplanets with those obtained by \cite{papa2005} for cores with masses $\sim 10$~M$_{\earth}$. In particular, they find a qualitatively similar decrease in the accretion rate with time to that which we have described in Section \ref{funcform}.  They do not plot accretion rate versus time explicitly.  Rather, they plot accretions rate versus total accreted mass.  In Figure \ref{fig:papacomparison}, we plot their results for 5 and 15 M$_{\earth}$ cores (dashed lines and dot-dashed lines respectively) and extrapolations of our 10~M$_{\earth}$ cases (solid lines).  In each case, the evolution of the accretion rates is given for standard (thick lines) and 1\% interstellar grain opacities (thin lines).  Since we have not been able to evolve our calculations until they had accreted gas with a mass comparable to the core mass, we have extrapolated our accretion rates from 20 orbits using the functional form described in Section \ref{funcform}.  Figure \ref{fig:papacomparison} shows that our accretion rates (for the 10~M$_{\earth}$ planetary core) are bracketed by their accretion rates obtained using planetary cores with slightly lower and higher masses.  We note that the decline of the accretion rates found by \citeauthor{papa2005} has the same form as we obtain until the accreted gas mass reaches approximately 1/4 of the core mass.  Beyond this point, the increasing gravitational potential of the protoplanet reverses the decline in the accretion rate.  We would expect our models to show a similar upturn if we were able to evolve them to a similar stage.

In summary, we find that our three-dimensional self-gravitating radiation hydrodynamical simulations give similar accretion rates to those of \cite{papa2005}, at least as far as we are able to evolve them.  The effect of reducing the grain opacity on the accretion rates of low-mass protoplanets is also similar to the results obtained from one-dimensional quasi-static models of protoplanet gas accretion.   In the future, we would like to evolve our calculations over longer periods to see whether the three-dimensional case displays significant differences to the one-dimensional quasi-static models. It might be that radiative transfer vertically through the protoplanetary disc would allow for more rapid cooling of a growing envelope, and consequently higher accretion rates than are found in the spherically-symmetric models.

\section{Conclusions}

We have investigated the effects of grain opacity, protoplanetary disc surface density, and protoplanet mass on the gas accretion of protoplanets immersed in a protoplanetary disc using self-gravitating radiation hydrodynamical calculations.  We have investigated protoplanets spanning the mass range 10-333 $\rm M_{\earth}$, with grain opacities ranging from interstellar values to 0.1\% of these values. 
For some of our models, we resolve the gas accretion process right down to the surface of the solid planetary core, the first time this has been done in three dimensional radiation hydrodynamical calculations.  The majority of our models resolve the gas flow down to 1\% of the Hill radius of the protoplanet.

We find that the inclusion of radiative transfer leads to lower protoplanetary accretion rates, the generation of weaker spiral shocks and less extended horseshoe orbit regions in the protoplanetary disc, and the prevention of circumplanetary disc formation for low-mass protoplanets compared with similar locally-isothermal calculations. Only cores with masses $\rm \gsim 100 M_{\earth}$ produce strong spiral shocks and form circumplanetary discs when using interstellar grain opacities. With reduced grain opacities the results shift towards the locally-isothermal results but the accretion rates are always lower than in the locally-isothermal calculations.  For high-mass protoplanets ($\gsim 100$~M$_{\earth}$) the accretion rates do not depend substantially on grain opacity, or in fact whether radiative transfer is employed at all, because gravity dominates over thermal effects. For lower-mass protoplanets modelled using core radii of 1\% of the Hill radius, the accretion rates can vary by up to 2 orders of magnitude, with lower opacities generally giving higher accretion rates.

For 33~M$_{\earth}$ cores, we also performed calculations with realistically-sized solid cores, modelling the gas flow down to the surface of the core.  These calculations were highly computationally intensive.  We found that the accretion rates did not change significantly for grain opacities of $10-100$\% of the interstellar opacities from those obtained using core radii of 1\% of the Hill radius.  However, for lower opacities of $0.1-1$\% of the interstellar values, smaller cores gave lower accretion rates.  Thus, we conclude that the enhancement of the accretion rates seen at very low opacities is probably not as large as indicated by the 1\% Hill radius calculations. It is therefore essential to model down to realistically-sized cores in order to obtain accurate accretion rates for low-mass cores at low opacities.

Finally, we compared our results to those obtained from one-dimensional quasi-static calculations of gas accretion by planetary cores.  Although we were unable to evolve our calculations until the mass accreted was a substantial fraction of the core mass, we found our accretion rates were in good agreement with the early evolution obtained from one-dimensional  models.  We also find that reducing the grain opacity leads to increased accretion rates in a quantitatively similar manner to  one-dimensional models.

\section*{Acknowledgments}

We acknowledge very incisive and useful comments by the anonymous referee. BAA and MRB thank Daniel Price and David Rundle for many useful conversations. The calculations reported here were performed using the University of Exeter's SGI Altix ICE 8200 supercomputer. MRB is grateful for the support of a Philip Leverhulme Prize and a EURYI Award. This work, conducted as part of the award ``The formation of stars and planets: Radiation hydrodynamical and magnetohydrodynamical simulations"  made under the European Heads of Research Councils and European Science Foundation EURYI (European Young Investigator) Awards scheme, was supported by funds from the Participating Organisations of EURYI and the EC Sixth Framework Programme. Many visualisations were produced using SPLASH \citep{splash}, a visualisation tool for SPH that is publicly available at http://www.astro.ex.ac.uk/people/dprice/splash.

\bibliography{paper.bib}

\end{document}